\documentclass{aa}

\usepackage{graphicx}
\usepackage{adjustbox}
\usepackage{txfonts}
\usepackage{url}
\usepackage{relsize}
\usepackage{array}
\usepackage{longtable}
\usepackage{supertabular}
\usepackage{setspace}

\usepackage{etoolbox}
\AtBeginEnvironment{longtable}{\singlespacing}

\newcolumntype{P}[1]{>{\centering\arraybackslash}p{#1}}
\newcolumntype{M}[1]{>{\centering\arraybackslash}m{#1}}

\begin{document}

\def\simlt{\mathrel{\rlap{\lower 3pt\hbox{$\sim$}}\raise 2.0pt\hbox{$<$}}}
\def\simgt{\mathrel{\rlap{\lower 3pt\hbox{$\sim$}} \raise 2.0pt\hbox{$>$}}}

\title{A search for candidate strongly-lensed dusty galaxies \\ in the \textit{Planck} satellite catalogues}


    \author{T.~Trombetti\inst{1}\fnmsep\thanks{e-mail: trombetti@ira.inaf.it}
                     \and
           C.~Burigana\inst{1,2,3}\fnmsep\thanks{e-mail: burigana@ira.inaf.it}
           \and
           M.~Bonato\inst{1,4,5}\fnmsep\thanks{e-mail: matteo.bonato@inaf.it}
           \and
           D.~Herranz\inst{6,7}\fnmsep\thanks{e-mail: herranz@ifca.unican.es}
           \and
           G.~De~Zotti\inst{5}\fnmsep\thanks{e-mail: gianfranco.dezotti@inaf.it}
           \and
           M.~Negrello\inst{8}\fnmsep\thanks{e-mail: NegrelloM@cardiff.ac.uk}
           \and
           V.~Galluzzi\inst{9}\fnmsep\thanks{e-mail: vincenzo.galluzzi@inaf.it}
           \and
           M.~Massardi\inst{1,4}\fnmsep\thanks{e-mail: massardi@ira.inaf.it}
           }

    \institute{INAF, Istituto di Radioastronomia, Via Piero Gobetti 101, I-40129 Bologna, Italy
    \and
    Dipartimento di Fisica e Scienze della Terra, Universit\`a di Ferrara, Via Giuseppe Saragat 1, I-44122 Ferrara, Italy
    \and
    INFN, Sezione di Bologna, Via Irnerio 46, I-40127 Bologna, Italy
    \and
    INAF, Italian ALMA Regional Centre, Via Piero Gobetti 101, I-40129, Bologna, Italy
    \and
    INAF, Osservatorio Astronomico di Padova, Vicolo dell'Osservatorio 5, I-35122 Padova, Italy
    \and
    Instituto de F\'{\i}sica de Cantabria, CSIC-UC, Av. de Los Castros s/n, E-39005 Santander,
    Spain
    \and
    Departamento de F\'{\i}sica Moderna, Universidad de Cantabria, 39005-Santander,
    Spain
    \and
    School of Physics and Astronomy, Cardiff University, The Parade, Cardiff CF24 3AA, UK
    \and
    Osservatorio Astronomico di Trieste, Via G. B. Tiepolo, 11 I-34143 Trieste, Italy
            }

   \date{Received ...; accepted ...}



  \abstract{The shallow, all-sky \textit{Planck} surveys at sub-millimeter wavelengths have detected the brightest strongly gravitationally lensed dusty galaxies in the sky. The combination of their extreme gravitational flux boosting and image stretching offers the unique possibility of measuring in extraordinary detail, via high-resolution imaging and spectroscopic follow-up, the galaxy structure and kinematics in early evolutionary phases, thus gaining otherwise unaccessible direct information on physical processes in action. However the extraction of candidate strongly lensed galaxies from \textit{Planck} catalogues is hindered by the fact that they are generally detected with poor signal-to-noise ratio, except for the few brightest ones. Thus their  photometric properties are strongly blurred so that they are very difficult to single out. We have devised a method capable of increasing by a factor of about three to four the number of identified \textit{Planck}-detected strongly lensed galaxies, although with an unavoidably limited efficiency. Our approach exploits the fact that strongly lensed galaxies have sub-millimeter colours definitely colder than nearby dusty galaxies that constitute the overwhelming majority of extragalactic sources detected by \textit{Planck}. The sub-mm colours of the 47 confirmed or very likely \textit{Planck}-detected strongly lensed galaxies have been used to estimate the colour range spanned by objects of this kind. Moreover, most nearby galaxies and radio sources can be picked up by cross-matching with the IRAS and PCNT catalogues, respectively. We present samples of lensed candidates selected at 545, 857 and 353\,GHz, comprising 177, 97 and 104 sources, respectively. The efficiency of our approach, tested exploiting data from the SPT survey covering $\simeq 2,500\,\hbox{deg}^2$, is estimated to be in the range 30\%--40\%. We also discuss stricter selection criteria increasing the estimated efficiency to $\simeq 50\%$ at the cost of a somewhat lower completeness. Our analysis of SPT data has identified a dozen of galaxies that can be reliably considered previously unrecognized \textit{Planck}--detected strongly lensed galaxies. Extrapolating the number of \textit{Planck}--detected confirmed or very likely strongly lensed galaxies found within the SPT and H-ATLAS survey areas, we expect from $\simeq 150$ to $\simeq 190$ such sources over the full $|b|>20^\circ$ sky.
 }

   \keywords{gravitational lensing: strong -- submillimeter: galaxies -- galaxies: high-redshift}

   \maketitle


\section{Introduction}\label{sect:intro}

An interesting and unexpected \citep[but see][]{Negrello2007} result from \textit{Planck} was the detection of ultra-bright strongly lensed high-$z$ sub-millimeter galaxies (SMGs)
discovered by Herschel with extreme magnifications, $\mu$, in
the range 10--50 \citep{Herranz2013, Canameras2015, Harrington2016, Harrington2020, DiazSanchez2017}. These objects offer a unique opportunity to get detailed information on the internal structure, gas properties and kinematics of high-$z$ galaxies during their most active, dust-enshrouded star-formation phase \citep{Fu2012, Nesvadba2016, Nesvadba2019, Canameras2017, Canameras2017ALMA, Canameras2018a, Canameras2018b, Canameras2020, Harrington2018, Harrington2019, Dannerbauer2019,  PlanckCollaboration2020BEEP}.

This information is absolutely crucial to understand the key processes
governing the galaxy formation and early evolution. Current galaxy formation
models envisage widely different physical mechanisms for shaping the galaxy
properties: mergers, interactions, cold flows from the intergalactic medium, in
situ processes \citep[for reviews see][]{SilkMamon2012, SomervilleDave2015}.
All models have a large number of adjustable parameters that allow them to be
consistent with the available statistical information (source counts, redshift
distributions).

The only way to get direct information on physical processes at work is to look
inside the high-$z$ star-forming galaxies. But these are compact, with typical
effective radii of 1--2\,kpc \citep[e.g.,][]{Shibuya2015, Shibuya2019, Spilker2016, Hodge2016, Ikarashi2017,
Enia2018, Fujimoto2018, Fujimoto2020}, corresponding to angular radii of 0.1--0.2 arcsec at $z\simeq 2$--3. Thus they are hardly resolved even by ALMA and by the HST. If they are resolved, high enough S/N
ratios per resolution element are achieved only for the brightest galaxies,
probably not representative of the general population.

Strong gravitational lensing provides a solution to these problems, allowing us
to study high-$z$ galaxies in extraordinary detail, otherwise beyond reach of
present-day instrumentation \citep[e.g.,][]{Sun2021}. This happens thanks to the magnification of the
galaxy flux combined with a stretching of images. Since lensing conserves the
surface brightness, the effective angular size is stretched on average by a
factor $\mu^{1/2}$.

A spectacular example are ALMA observations with a $0.1''$ resolution of the strongly
lensed galaxy PLCK\_G244.8\-+54.9 at $z \simeq 3.0$  with $\mu \simeq 30$
\citep{Canameras2017ALMA}: they reached the astounding spatial
resolution of $\simeq 60\,$pc, substantially smaller than the size of Galactic
giant molecular clouds. \citet{Canameras2017ALMA} have also obtained CO
spectroscopy, measuring the kinematics of the molecular gas with an uncertainty
of 40--50 km/s. This spectral resolution makes possible a direct investigation
of massive outflows driven by AGN feedback at high $z$, with predicted
velocities of $\sim 1000\,\hbox{km}\,\hbox{s}^{-1}$ \citep{KingPounds2015}.

Outflows are advocated by all the main galaxy formation models to explain the
star-formation inefficiency in galaxies (only $\sim 10\%$ of baryons end up in
stars). However the observational confirmation of outflows of the direct fuel for star formation (namely, molecular gas) is very difficult to achieve at high-$z$ due to the weakness of their spectral signatures \citep[for a review see][]{Veilleux2020}. Even when outflows are detected, a proper assessment of their properties is limited by spatial resolution and sensitivity of instruments.

Strong lensing allowed \citet{Spilker2018} and \citet{Jones2019} to detect, by means of ALMA spectroscopy, fast, massive molecular outflows in galaxies at $z= 5.293$ and 5.656, respectively, discovered by the South Pole Telescope (SPT) survey. \citet{Spilker2020} found unambiguous evidence for outflows in 8 out of 11 SPT lensed galaxies at $z>4$.

\begin{table*}
\caption{Confirmed strongly lensed galaxies detected by \textit{Planck}. Flux densities are in mJy. They are taken from the PCCS2 catalogue except for the source at RA=36.6416\,deg, DEC=23.7579\,deg which is in the PCCS2E catalogue (sources of unknown reliability) and for that at RA=143.0985\,deg, DEC=27.4167\,deg, which is one of the \textit{Planck} high-$z$ source candidates \citep[PHz catalogue;][]{PlanckCollaboration2016highz}. The redshifts of the source and of the lens are denoted by $z$ and $z_l$, respectively. In the first column, P stands for PCCS2, PE for PCCS2E; the adjacent number is the selection frequency (GHz).}
\label{tab:confirmed}
\begin{center}
\begin{tabular}{lrrcccccc}
\hline
  \multicolumn{1}{c}{\textit{Planck} name}  &   \multicolumn{1}{c}{RA (deg)}  &   \multicolumn{1}{c}{DEC (deg)}  &    \multicolumn{1}{c}{$S_{353}$} &   \multicolumn{1}{c}{$S_{545}$} &   \multicolumn{1}{c}{$S_{857}$} &   \multicolumn{1}{c}{$z$}  &   \multicolumn{1}{c}{$z_l$}  &  \multicolumn{1}{c}{Ref}  \\
\hline
P353\,G293.74-69.76 & 17.4580  & -47.0360 & $273\pm 49$ &   --         &  $741\pm 132$ & 3.614  & 0.669 & S16, R20 \\
P545\,G190.40-83.77 & 19.1949  & -24.6172 &    --       & $684\pm 102$ & $1138\pm 132$ & 2.1245 & 0.4 &  H20 \\
P545\,G287.12-68.67 & 21.2800  & -47.3987 & $226\pm 46$ & $517\pm 79$  & $903\pm 113$  & 2.5148  & 0.305 &  W13 \\
P545\,G160.59-56.77 & 32.4210  &   0.2626 & $329\pm 66$ & $813\pm 109$ & $1309\pm 165$ & 2.5534  & 0.202 & H16 \\
PE857\,G149.41-34.16 & 36.6416  & 23.7579  & $498\pm 78$ & $1174\pm 126$ & $2263\pm 250$ & 3.1190 & 0.34 & H20 \\ 
P857\,G227.77-60.61 & 46.2943  &-30.6084  & --          &    --         & $613\pm 101$  & 2.2624 & 0.1--0.5 & H20 \\
P545\,G157.43+30.34 & 117.2155 & 59.6982  & $407\pm 67$ &  $927\pm 108$ & --            & 2.7544 & 0.402 & H20 \\
P545\,G211.62+32.22 & 131.7090 & 15.0965  & $431\pm 58$ &  $1114\pm 89$ & $1660\pm 117$ & 2.6615 & 0.1 & H20 \\
PHz\,G200.61+46.09 & 143.0985 & 27.4167  & $370\pm 190$ & $586\pm 68$ & $747\pm 78$ & $\simeq 3.0$ & 0.6 & C15\\ 
P545\,G145.25+50.84& 163.3439 &  60.8635 & $450\pm  50$ & $782\pm 90$ &   --        & 3.6000 & --  & C15\\
P545\,G244.76+54.94 & 163.4710 & 5.9392   & $444\pm 49$ & $915 \pm 91$ & $1334\pm 131$ & 3.0055 & 1.525 & C15, H20\\
P857\,G158.56+64.72 & 171.8060 & 46.1567  &  --         &  --          & $886\pm 128$ & 1.3036  & 0.415 & H20\\
P857\,G188.25+73.11 & 174.5230 & 32.9658  &  --         &  --          &  $882\pm 145$ & 2.0183 &  0.6 & H20\\
P545\,G231.27+72.22 & 174.8406 & 20.4147  &  --         & $533\pm 85$  &  --           & 2.8584 & 0.57 & C15, H20\\
P353\,G270.57+58.50 & 176.6579 & -0.1922  & $287\pm 50$ &  --          & $956\pm 148$   & 3.2592  & 1.2247 & F12, He13, N17\\
P545\,G138.59+62.02 & 180.5320 & 53.5778  &  --         & $633\pm 77$  & $835\pm 106$  & 2.4416 & 0.212  &  H16 \\
P353\,G076.26+79.96 & 201.6254 & 33.7353  & $255\pm 49$ & --           &    --         & 2.9507 & 0.7856  &  N17 \\
P545\,G007.97+80.28 & 202.3920 & 22.7242  & $283\pm 56$ & $921\pm 97$  & $1612\pm 121$ & 2.0401 & 0.443 & D17 \\
P545\,G104.43+66.26 & 204.1456 & 49.2204  &   --        & $539\pm 80$  & $852\pm 126$  & 3.2548 & 0.28 & H20\\
P857\,G052.27+77.90 & 206.1225 & 30.5094  &   --        &    --        & $705\pm 130$  & 2.3010 & 0.6721 & N17\\
P857\,G030.03+62.79 & 222.4941 & 22.6436  &   --        &    --        & $703\pm 125$ &  2.1536 &-- & H20\\
P545\,G045.11+61.10 & 225.6502 & 29.3475  & $333\pm 64$ & $498\pm 86$  &  --          &  3.4270 & 0.56 & C15, Ne16 \\
P545\,G080.25+49.86 & 236.1350 & 50.3961  &   --        & $504\pm 77$  &  --          &  2.5988 & 0.673 & C15, H20\\
P857\,G107.64+36.93 & 241.8242 & 73.7842  &   --        &  --          & $761\pm 104$ &  1.4839 &  0.65 & H16, H20 \\
P545\,G092.49+42.89 & 242.3240 & 60.7558  & $307\pm 45$ & $788\pm 79$  & $1240\pm 113$ & 3.2555  & 0.45 & H16, H20\\
P545\,G053.44-36.27 & 323.7983 & -1.0478  & $337\pm57$  & $775\pm 109$ & $947\pm 165$ & 2.3259   & 0.325 & Sw10\\                         	             	
P353\,G325.97-59.46 & 353.0972 & -53.9804 & $307\pm 52$ &    --        &  --          &  2.73   &  --    & Su21\\
\hline
\end{tabular}
\end{center}
\bigskip
\textbf{References.}
C15, \citet{Canameras2015}; D17, \citet{DiazSanchez2017};  F12, \citet{Fu2012}; H16, \citet{Harrington2016}; H20, \citet{Harrington2020}; He13, \citet{Herranz2013}; N17, \citet{Negrello2017lensed};  Ne16, \citet{Nesvadba2016};  R20, \citet{Reuter2020}; S16, \citet{Spilker2016}; Su21, \citet{Sun2021}; Sw10, \citet{Swinbank2010};W13, \citet{Weiss2013}.
\end{table*}

\begin{figure}[tbh]
\center
\includegraphics[width=0.49\textwidth]{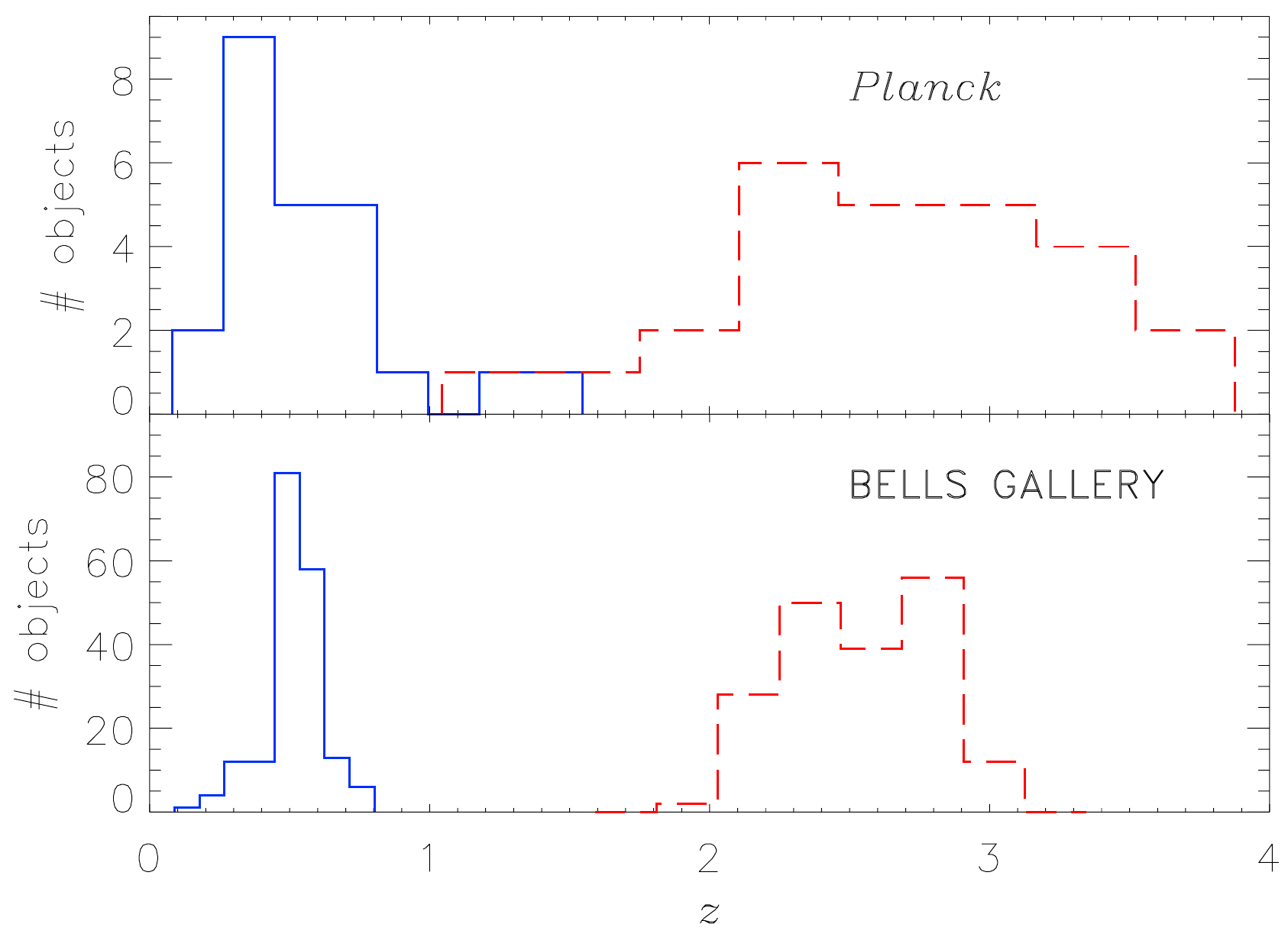}
\caption{Redshift distributions of the background lensed galaxies (solid blue histograms) and of the foreground lenses (dashed red histograms) detected by \textit{Planck} compared with those of the 187 strong gravitational lens candidates in the BELLS GALLERY survey parent sample  \citep{Shu2016}.}\label{fig:PlanckSDSS}
\end{figure}

Prolific searches of strongly lensed galaxies have been carried out in the optical \citep[][and references therein]{Cao2020, Talbot2020}. The mm/sub-mm surveys not only complement the optical ones by extending the selection to dust-enshrouded galaxies but also reach higher redshifts both of background lensed galaxies and of foreground lenses, as illustrated by Fig.~\ref{fig:PlanckSDSS}. In fact, the mm/sub-mm region is exceptionally well suited to reach high redshifts due to the large, negative K-correction and to the strong cosmological evolution. A further advantage is that lensed SMGs are generally free from blending with foreground lenses, showing up in different wavebands.

\citet{Negrello2007} predicted that essentially all high-$z$ galaxies brighter than $S_{500\,\mu\rm m}= 100\,$mJy detected by \textit{Herschel} surveys would have been strongly lensed (magnification $\mu \ge 2$) and pointed out that they  could  be identified with close to 100\% efficiency since the other extragalactic sources above that flux density limit would have been easily recognizable local galaxies plus a small fraction of radio sources. This prediction proved to be accurate and a total of about 170 candidate strongly lensed galaxies with $S_{500\,\mu\rm m}\ge 100\,$mJy have been selected by \citet{Negrello2010, Negrello2017lensed}, \citet{Wardlow2013} and \citet{Nayyeri2016} respectively from the \textit{Herschel} Astrophysical Terahertz Large Area Survey \citep[H-ATLAS;][]{Eales2010}, the \textit{Herschel} Multitiered Extragalactic Survey \citep[HerMES;][]{Oliver2012} and the HerMES Large Mode Survey (HeLMS) plus \textit{Herschel} Stripe 82 Survey \citep[HerS;][]{Viero2014} catalogues.

To enlarge the sample of \textit{Herschel}--detected strongly lensed galaxies it is necessary to go to lower flux densities, where these objects are mixed up with an increasing fraction of unlensed high-$z$ galaxies. Methods proposed to extract strongly lensed galaxies exploit the fact that they are located within arc-seconds from the galaxy acting as the lens \citep{GonzalezNuevo2012, GonzalezNuevo2019, Bakx2020}. The latter galaxies are close enough to the \textit{Herschel} galaxies to be interpreted as their optical counterparts by likelihood ratio techniques \citep[e.g.,][]{Bourne2016} but cannot be the sources themselves because they generally are massive ellipticals containing old stellar populations, hence with negligible far-IR/sub-mm emission.

\textit{Planck} sub-mm surveys are much shallower that the \textit{Herschel} ones: their detection limits are more than one order of magnitude higher. However, thanks to its all-sky coverage the \textit{Planck} mission had the unique
capability of detecting the brightest strongly lensed high-$z$ SMGs in the sky, i.e. those best suited to get high spatial and spectral resolution follow-up data. In fact, the shallowness of \textit{Planck} surveys implies
that only really extreme magnifications can boost high-$z$ galaxy flux
densities above the detection limits. To put the argument in context, let us
remember that essentially all high-$z$ SMGs brighter than 100\,mJy at 600\,GHz
($500\,\mu$m) were found to be strongly lensed \citep{Negrello2010, Negrello2017lensed, Wardlow2013, Nayyeri2016}, but the brightest candidate strongly lensed galaxy detected over the $602\,\hbox{deg}^2$ of the
\textit{Herschel} Astrophysical Terahertz Large Area Survey \citep[H-ATLAS;][]{Eales2010} has a flux density of 465.7\,mJy at 857\,GHz \citep{Negrello2017lensed}. For comparison, the 90\% completeness limit at this frequency of the second \textit{Planck} Catalogue of Compact Sources \citep[PCCS2;][]{PCCS2} in the ``extragalactic zone'' is of 791\,mJy. Correspondingly, the estimated gravitational magnifications of \textit{Planck}-detected lensed galaxies, mostly in the range 10 to 50 \citep{Canameras2015, Harrington2020}, are substantially higher than those of H-ATLAS lensed galaxies, which are in the range $\sim 5$--15 \citep{Negrello2017lensed, Enia2018}. The unique possibilities offered by the \textit{Planck} selection are clear.

Searching the literature we have collected a total of 27 \textit{Planck}-detected strongly lensed galaxies  (Table~\ref{tab:confirmed}). All but two of them are listed in the Second \textit{Planck} Catalogue of Compact Sources \citep[PCCS2;][]{PCCS2}, one in the PCCS2E which contains sources with unknown reliability,  and one in the catalogue of \textit{Planck} high-$z$ source candidates, detected in the cleanest 26\% of the sky \citep[PHz catalogue;][]{PlanckCollaboration2016highz}.

\citet{Canameras2015} list 3 more sources as ``\textit{Planck}--detected'' (at RA, DEC in deg: 139.619, 51.7064; 171.8108, 42.4736; 217.3249, 59.3525). Other 4 sources are listed by \citet{Harrington2016} and/or \citet{Harrington2020} at: 200.5730, 9.3907; 200.7615, 55.6003; 217.0995, 35.4389; 348.4860, 1.1549. None of them appears in the last versions of the online \text{Planck} catalogues.

Measured redshifts are in the range 1.3--3.6, implying that a statistically well defined sample of \textit{Planck} lensed galaxies would be very well suited to investigate the structure of galaxies across
the peak of the cosmic star formation rate. However so far searches have been fragmentary and over limited sky areas, implying that the available sample is highly inhomogeneous and incomplete.

Finding the rare \textit{Planck}--detected strongly lensed galaxies among the several thousands local dusty galaxies is not an easy task, however. Statistical selection techniques such as those successfully used to select \textit{Herschel}--detected strongly lensed candidates fainter than $S_{500\,\mu\rm m}=100\,$mJy \citep{GonzalezNuevo2012, GonzalezNuevo2019, Bakx2020}, mentioned above, cannot be applied: the typical rms positional error of the relevant sources, computed using eq.~(7) of \citet{PCCS2}, is of 1.5 arcmin, i.e. a factor of 26 larger than the typical positional error for \textit{Herschel}--SPIRE sources \citep[3.4 arcsec;][]{Bourne2016}.

To look for counterparts to H-ATLAS sources, \citet{Bourne2016} considered SDSS galaxies with $r_{\rm model}<22.4$. Their surface density is $\simeq 1.2\times 10^4\,\hbox{deg}^{-2}$ so that
within the $2\,\sigma$ search radius of 3 arcmin there are, on average, $\simeq 94$ SDSS galaxies. Similarly, the surface density of WISE galaxies is $\simeq 9.9\times 10^3\,\hbox{deg}^{-2}$ \citep{Jarrett2017} so that there are, on average, $\simeq 80$ WISE galaxies within a 3 arcmin radius. For comparison, the number of chance associations within $10''$,  the search radius generally used in the \textit{Herschel} case, is $\simeq 0.09$ (SDSS) or $\simeq 0.08$ (WISE). We could not find any valid criterion to identify plausible foreground lenses among the many tens of galaxies lying within the \textit{Planck} search radius.

Another possibility would be to use Machine Learning techniques to select lensed candidates. This would require massive simulations using, e.g. the \textit{Planck} Sky Model \citep{Delabrouille2013} to produce a tailored training set. However, we have preferred to resort to the work-intensive approach described in Sect.~\ref{sect:method} where we also present our results. In Sect.~\ref{sect:conclusions} we summarize and discuss our main conclusions.

\begin{figure}
\includegraphics[width=0.98\columnwidth]{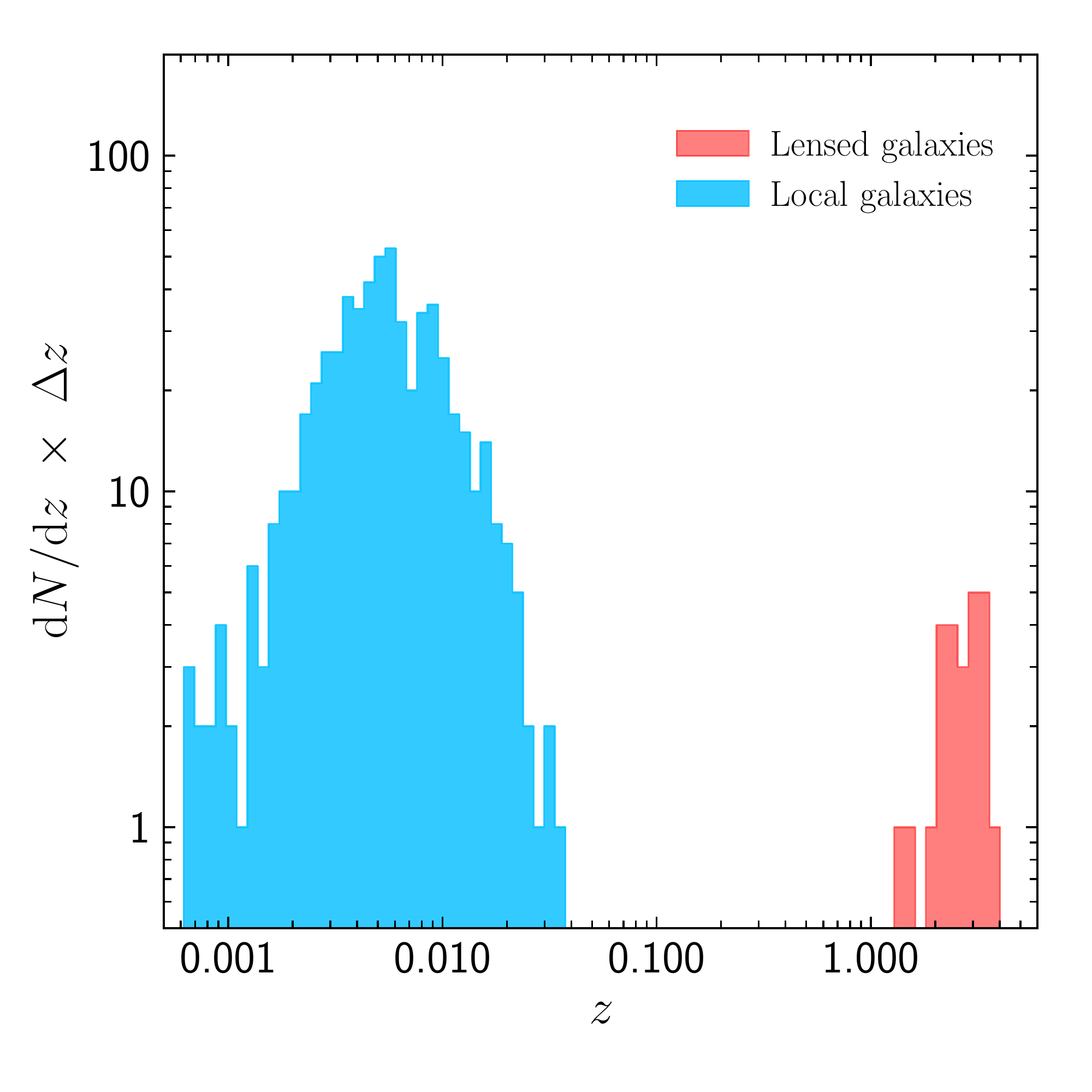}  
\caption{Redshift distribution of \textit{Planck} Early Release Compact Source Catalogue \citep[ERCSC;][]{ERCSC2011} galaxies detected at 545\,GHz \citep[][no more recent redshift distribution for a complete sample of \textit{Planck} dusty galaxies is available]{Negrello2013}. They are all local ($z<0.1$). At fainter flux densities, strongly lensed galaxies at much higher redshifts ($z\simgt 1$) begin to appear, with nothing in between. Plotted here is the redshift distribution of known strongly lensed galaxies detected by \textit{Planck} (Table~\ref{tab:confirmed}). The broad gap among the two populations persists at least down to a flux density of 100\,mJy at $600\,$GHz \citep{Negrello2017lensed}.    }
\label{fig:zdistr}
\end{figure}

\begin{figure}[tbh]
\center
\includegraphics[width=0.5\textwidth]{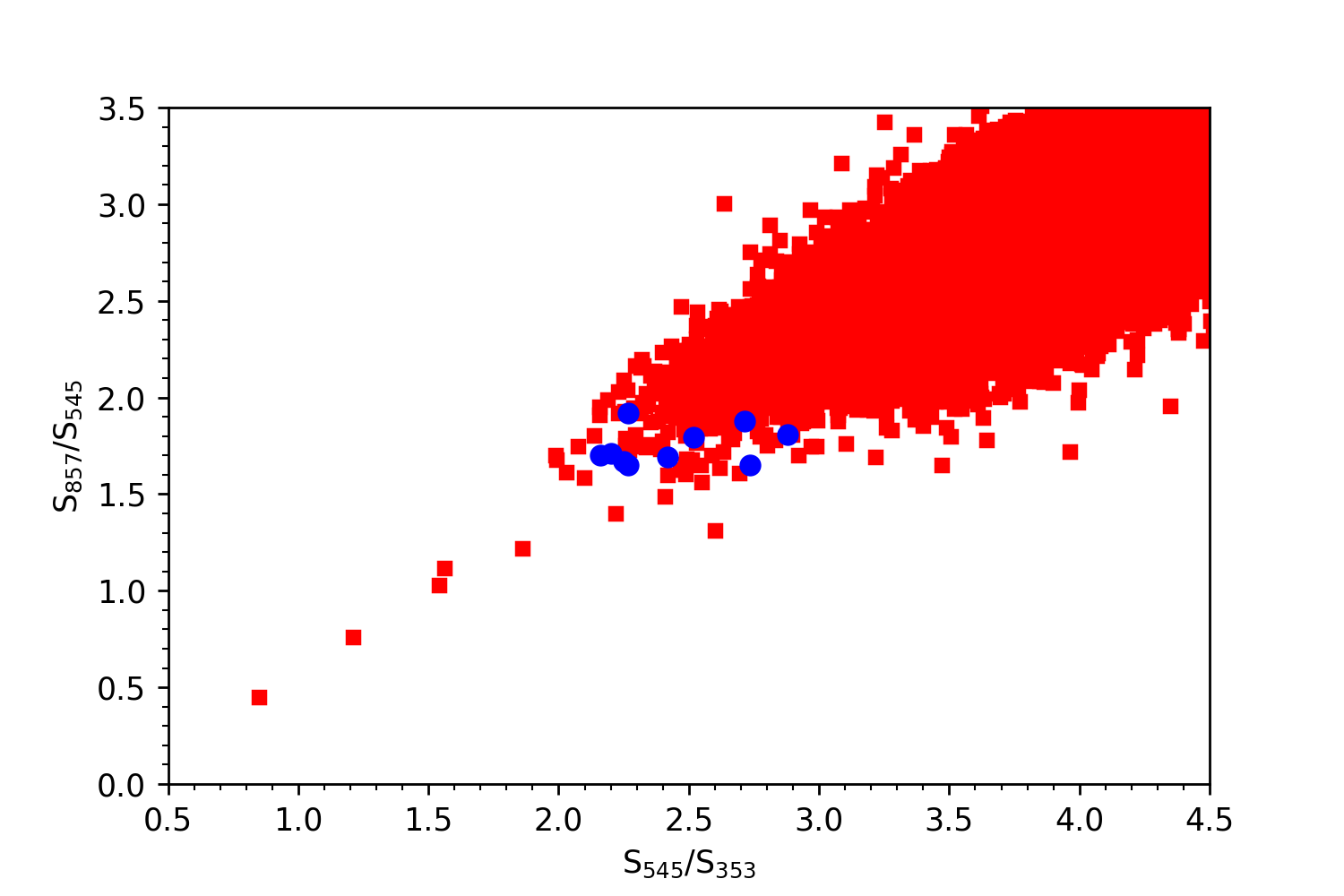}
\caption{Distribution  in the $S_{857}/S_{545}$ vs $S_{545}/S_{353}$ colour--colour plot of \textit{Planck}-detected galaxies with $S_{545}\ge 500\,$mJy. For this diagram we have used the multi-band BeeP photometry  \citep{PlanckCollaboration2020BEEP}.  Radio sources were removed by cross-matching the sample with the PCNT catalogue \citep{PlanckCollaboration2018PCNT}. The 10 confirmed strongly lensed galaxies in this sample (filled blue circles) have colours at the red end of the distribution of the other galaxies (filled red squares). Still redder objects are likely Galactic cold clumps (see text).}\label{fig:colours}
\end{figure}

\begin{figure*}
\includegraphics[width=\columnwidth]{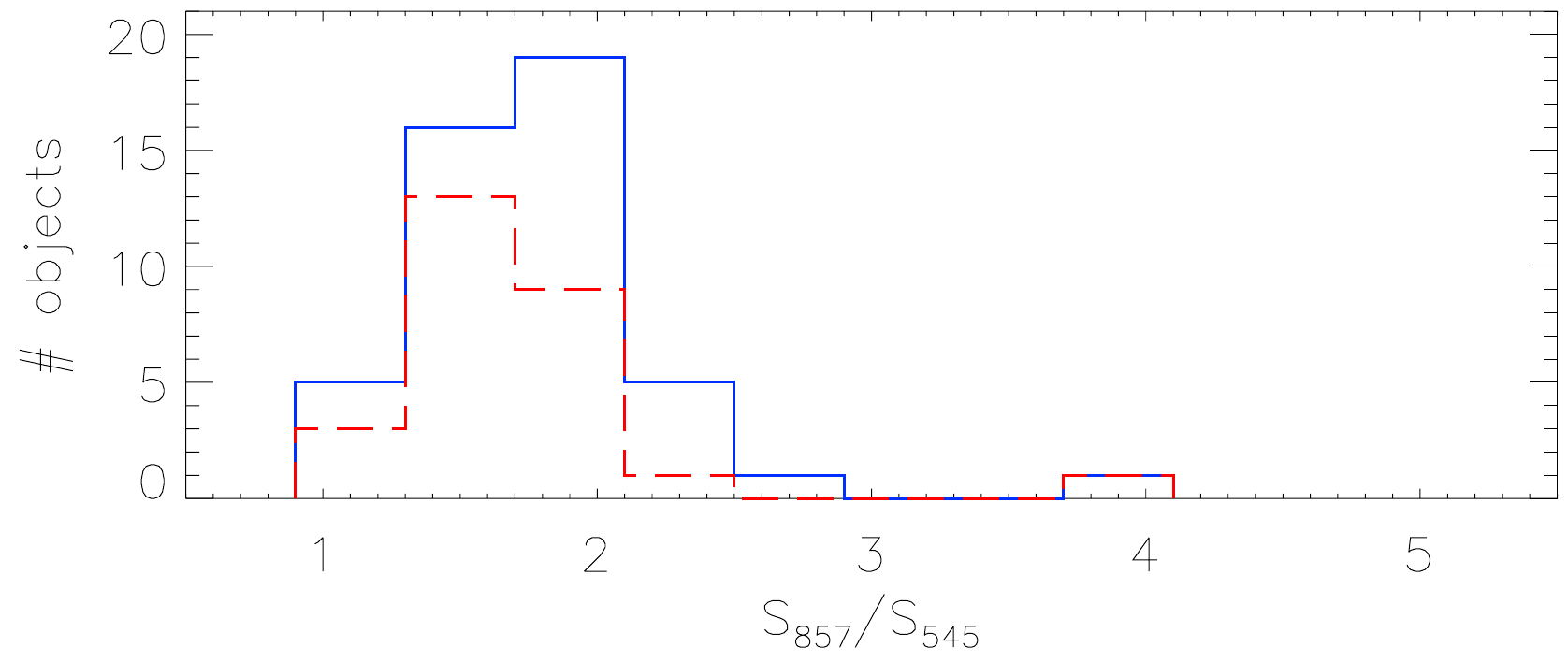}   
\includegraphics[width=\columnwidth]{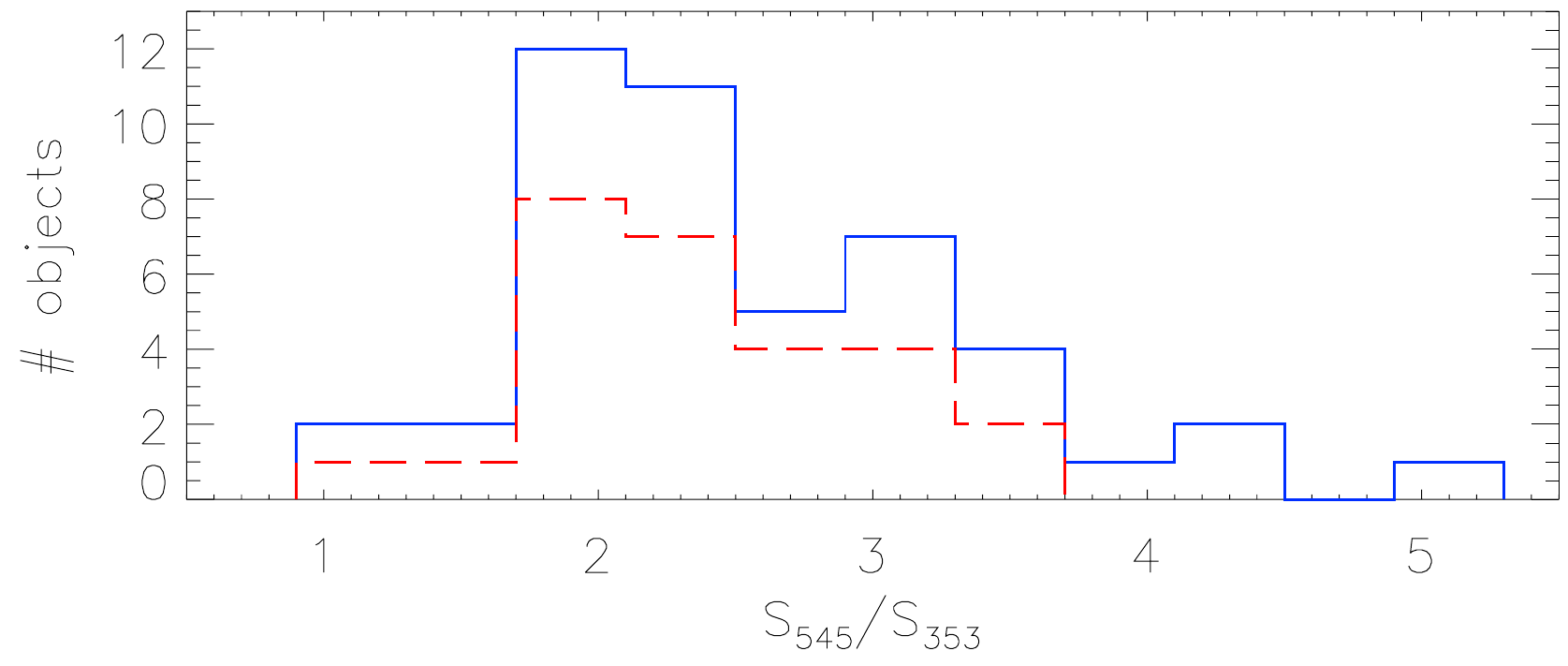}   
\caption{Distributions of sub-mm colours $S_{857}/S_{545}$ and $S_{545}/S_{353}$ of the 47 confirmed plus very likely \textit{Planck} detected strongly lensed galaxies solid blue histograms. The dashed red histograms show the distributions of the 27 confirmed strongly lensed galaxies only.}
\label{fig:hist_lensed}
\end{figure*}

\section{Method}\label{sect:method}

\subsection{Overview}

As mentioned above, picking up strongly lensed galaxies from \textit{Planck} catalogues is not easy
since they are a tiny fraction of detected sources and their flux densities are
generally near the detection limit, as expected given the steepness of the bright end of their source counts \citep{Perrotta2002, Perrotta2003, Negrello2007, Vieira2010, Mocanu2013, Negrello2017lensed, Everett2020}. Sub-mm PCCS2 sources at high Galactic latitude ($|b|>20^\circ$)\footnote{At low Galactic latitudes the reliability of source detection cannot be accurately assessed because of the confusion from Galactic cirrus emission. Therefore the \textit{Planck} Collaboration has adopted a set of Galactic masks, defined by \citet{PlanckCollaboration2014}, to exclude regions to various levels of dust contamination. Our choice, $|b|>20^\circ$, roughly corresponds to the region outside the \textit{Planck}  G65 mask. } are mostly nearby star-forming galaxies with a small fraction of extragalactic radio sources which dominate at cm and mm
wavelengths \citep{PlanckCollaboration2013counts}. In addition there are galaxy
over-densities \citep{PlanckCollaboration2015clumpsHerschel, PlanckCollaboration2016highz}, Galactic cirrus \citep{Herranz2013} and Galactic cold clumps \citep[GCC;][]{PlanckCollaboration2016cold_cores}, intensity peaks of the cosmic infrared background (CIB) plus the rare strongly lensed galaxies we are looking for.

Since we are interested in dusty galaxies, we have considered only \textit{Planck} channels at $\nu \ge 353\,$GHz. As predicted by \citet{Negrello2007} and confirmed by the analysis of H-ATLAS data \citep{Negrello2017lensed}, at the bright detection limits of the \textit{Planck} sub-mm surveys, unlensed dusty galaxies are at $z\simlt 0.1$ \citep[see also][]{Negrello2013} while lensed galaxies are at $z> 1$ (see Fig.~\ref{fig:zdistr}) and therefore have substantially colder sub-mm colours.

Thus sub-mm colours are a distinctive property of strongly lensed galaxies. But can these sources be selected simply based on colours? The PCCS2 photometry is frequently limited to one or two frequencies, not enough to answer this question. Fortunately multi-band photometry obtained with the Bayesian Extraction and Estimation Package (BeeP) has been recently published \citep{PlanckCollaboration2020BEEP}.  We selected sources with $S_{545}\ge 500\,$mJy at high Galactic latitudes ($|b|\ge 20^\circ$). This flux density limit is slightly lower than the 90\% completeness limit in the ``extragalactic zone'' given by \citet[][]{PCCS2}, 555\,mJy, but consistent with the limit obtained from the comparison between \textit{Planck} and H-ATLAS photometry \citep{Maddox2018}.

Radio sources were identified and removed by cross-matching the BeeP sample with the \textit{Planck} multi-frequency catalogue of non-thermal sources \citep[PCNT;][]{PlanckCollaboration2018PCNT}. The distribution of sources in the $S_{857}/S_{545}$ vs $S_{545}/S_{353}$ plane is shown in Fig.~\ref{fig:colours}. The 10 confirmed strongly lensed galaxies comprised in the BeeP sample populate the red end of the distribution but their positions in the diagram are not clearly separated from those of unlensed galaxies. Hence, colours are useful to remove most nearby galaxies but still these are by far more numerous than lensed galaxies in any region encompassing the colours of the latter. The reason is that the uncertainties in the flux densities measured by \textit{Planck} are so large that the differences in colour between local and lensed galaxies are blurred. Thus the selection must be refined.

Figure~\ref{fig:colours} also shows that there are objects with colours even redder than those of strongly lensed galaxies. Two out of the 5 objects with the reddest colours (those at RA=83.9031\,deg, DEC=22.012\,deg; RA=142.4874\,deg, DEC=-23.2730\,deg, with $S_{857}/S_{545}=0.76$, $S_{545}/S_{353}=1.21$ and $S_{857}/S_{545}=1.12$, $S_{545}/S_{353}=1.56$, respectively) are listed in the \textit{Planck} catalogue of Galactic Cold Clumps \citep{PlanckCollaboration2016cold_cores}. The other 3 (170.8088, -48.6199; 277.4627, 1.4644; 343.5329, 16.1255 with $S_{857}/S_{545}=1.22$, $S_{545}/S_{353}=1.86$;  $S_{857}/S_{545}=1.03$, $S_{545}/S_{353}=1.54$ and $S_{857}/S_{545}=0.45$, $S_{545}/S_{353}=0.84$, respectively) are also likely Galactic cold clumps, although the data do not allow a firm classification yet.

The BeeP catalogue contains only sources listed in the PCCS2$+$2E at 857 GHz. It misses some of the reddest sources showing up at lower frequencies, including most of the known \textit{Planck}--detected strongly lensed galaxies. Thus, to achieve a comprehensive selection of strongly lensed candidates we need to go back to the PCCS2 catalogues at 353 and 545\,GHz. Again, we confined ourselves to sources at $|b|\ge 20^\circ$ but, to be as inclusive as possible, we did not impose any flux-density cut. We adopted the \textit{Planck} DETFLUX photometry because of its higher sensitivity compared to APERFLUX, although the latter is more robust at $\ge 353\,$GHz \citep{PCCS2}. In fact, using DETFLUX our selection criteria, specified in the sub-sections below, recover more confirmed strongly lensed galaxies than using APERFLUX. For example, at 545\,GHz we recover 18 confirmed strongly lensed galaxies, while only 11 are recovered using APERFLUX. This happens because APERFLUX yields substantially lower signal-to-noise ratios for our faint sources. Moreover, their colours are spread over the region occupied by local dusty galaxies. Hence selection criteria based on colours are much less efficient:  going deeper is crucial for our purposes.

A first cleaning of the initial, $|b|\ge 20^\circ$, samples at each frequency was obtained cross-matching them with the IRAS PSC/FSC Combined Catalogue \citep{Abrahamyan2015} using a 3\,arcmin search radius. Since the dust emission spectrum of low-$z$ galaxies generally peaks in the range 60--$150\,\mu$m \citep{Lagache2005}, i.e. within or close to the IRAS wavelength range, IRAS is substantially more sensitive than \textit{Planck} to these objects. Not all IRAS galaxies are at low $z$. About 4\% are at $z>0.3$ and a small fraction (0.7\%) are hyper-luminous infrared galaxies and dusty QSOs at $z$ of up to $\simeq 4$, including 4 strongly lensed galaxies \citep{RowanRobinson2018}. However, none of the IRAS galaxies with extreme IR
luminosities, listed in Table~5 of \citet{RowanRobinson2018}, has a PCCS2 counterpart. Thus dropping PCCS2 sources with IRAS counterparts we rid the sample of nearby dusty galaxies without affecting high-$z$ objects.

The removal of radio sources is slightly less straightforward. Matches with the PCNT include dusty galaxies hosting radio nuclei. As far as their sub-mm colours are dominated by dust emission, they should be dealt with as the other dusty galaxies. We have therefore inspected the matches one by one checking whether, after subtracting the radio contribution extrapolated from lower frequencies, the sub-mm colours were consistent with dust emission, i.e. the continuum spectra showed a steepening from mm to sub-mm wavelengths. Objects with spectra consistent with being dust-dominated at sub-mm wavelengths were kept.  At first we thought that also sources with counterparts in the \textit{Planck} GCC catalogue should be removed, but we gave up on that because we found that there are 8 confirmed strongly lensed galaxies among GCC's. This is not really surprising. Although \citet{PlanckCollaboration2016cold_cores} applied three independent methods to remove extragalactic sources from their sample, they didn't have any way to identify strongly lensed galaxies which have colours similar to GCC's.

To clean the samples further we made a selection based on sub-mm colours. The obvious benchmark for this purpose is the sample of confirmed strongly lensed \textit{Planck}--detected galaxies (Table~\ref{tab:confirmed}). Since the number of these objects is limited, we have complemented it with other \textit{Planck} galaxies which have properties indicating that they are very likely strongly lensed.

Negrello et al. (2021, in preparation) carried out SCUBA\,2 observations of a preliminary sample of candidate strongly lensed galaxies with $S_{545}\ge 500\,$mJy, detecting 12 of them. The SCUBA\,2 detection implies that these objects are point like, i.e. not cold extended objects like cold clumps, proto-clusters of high-$z$ dusty galaxies or positive fluctuations of the CIB. Their red colours imply substantial redshifts, but galaxies at substantial redshifts are almost certainly strongly lensed. This point is illustrated by Fig.~\ref{fig:zdistr} showing a striking bimodality of the redshift distribution of \textit{Planck}--detected galaxies. On one side we have nearby late-type galaxies, at $z\simlt 0.1$, and hence easily recognizable in optical/near-infrared catalogues. On the other side we have dust enshrouded, hence optically very faint, gravitationally lensed galaxies at $z\simgt 1$. The bimodality is inherent in shallow sub-mm surveys and was shown to persist down to detection limits much deeper than \textit{Planck}'s; it is seen in \textit{Herschel} surveys for $S_{\rm lim, 600}=100\,$mJy \citep{Negrello2017lensed}. \textit{Planck}--detected galaxies at $z>0.2$ would be hyperluminous infrared galaxies (HyLIRGs, $L_{\rm IR}> 10^{13}\,L_\odot$). HyLIRGs are not detected at redshifts of a few tenths \citep{Gruppioni2013}; if they were present, they would have been detected by IRAS \citep{RowanRobinson2018}. On the other hand, the probability of SMGs undergoing strong lensing is heavily suppressed at $z\simlt 1$ \citep{Perrotta2002, Negrello2007, HezavehHolder2011}. Only at higher redshifts there are enough very luminous IR galaxies and the optical depth for strong gravitational lensing is large enough to yield the extreme amplifications needed to make galaxies detectable by \textit{Planck} (to reach the \textit{Planck} detection limits, galaxies must be both intrinsically ultraluminous and very highly magnified). Three of the 12 sources detected by Negrello et al. (2021) with SCUBA\,2 were later confirmed by \citet{Harrington2020} to be strongly lensed.

Furthermore, a cross-match of galaxies detected by \textit{Planck} with the catalogues of SPT galaxies \citep{Everett2020} yielded 4 more sources with redshifts indicative of strong lensing. Finally we added the 7 galaxies listed as \textit{Planck}--detected by either \citet{Canameras2015} or \citet{Harrington2016, Harrington2020}, not listed in the PCCS2 but present in earlier versions of the \textit{Planck} point source catalogues.

Most of these 47 sources don't have PCCS2 photometry at all 3 frequencies of interest (353, 545 and 857 GHz), as necessary to determine their sub-mm colours. To get uniform photometry for the full sample we performed a multi-frequency analysis with the ``Matrix multi-Filter'' methodology described by \citet{PlanckCollaboration2018PCNT}. This technique allowed us to get flux density estimates or upper limits at all frequencies of interest and also to improve the signal-to-noise ratio at the frequencies for which PCCS2 flux densities are available.

We note, in passing, that there are significant differences between the BeeP and the MTXF photometry.  The BeeP photometry is based on the all-sky temperature maps at 353, 545, and 857\,GHz from the \textit{Planck} 2015 release \citep[PR2;][]{PlanckCollaboration2016data_products} which was also the source for the PCCS2 photometry.  The Beep catalogue provides two sets of flux-density estimates, based on different models for the spectral energy distribution: the Modified Blackbody (MBB) and the Free model. The two sets are in good agreement with each other. We have chosen the MBB flux densities which seem to benefit from a better background subtraction.

The MTXF photometry, instead, exploits the most recent publicly available release in the \textit{Planck} Legacy Archive \citep[PR3;][]{PlanckCollaboration2020overview}. However we have checked that using PR2 maps we get really minor differences, $\simeq 2\%$. Much larger differences are produced by the different methods for measuring source flux densities. The BeeP photometry is in good agreement with the PCCS2 APERFLUX but not with DETFLUX. For example, at 857\,GHz the DETFLUX values are on average $\sim 24\%$ lower than the BeeP ones. Hence the BeeP photometry is close to aperture photometry.

In its current implementation, the MTXF method assumes instead that sources are point-like, i.e. that their spatial profile is that of the instrumental beam. It is therefore closer to DETFLUX and works well for the high-$z$ sources we are interested in. On the other hand, it is bound to underestimate the flux density of extended sources, such as nearby dusty galaxies. This is indeed what we see. The ratio between MTXF and BeeP flux densities is lower for the brightest sources which are very nearby dusty galaxies.  The mean MTXF/BeeP ratios  for the 388 common sources in the parent sample are 0.78, 0.75 and 0.84 at 857, 545 and 353\,GHz, respectively, close to the mean ratios between DETFLUX and BeeP flux densities. However, the ratios approach unity if we restrict ourselves to the weaker sources of interest here.

Another factor that may affect flux density estimates is positional accuracy. There are some differences between the positions where MTXF and BeeP locate the sources. \citet{PlanckCollaboration2020BEEP}  argue that positional offsets can account for up to a 5\% differences between BeeP and PCSS2 APERFLUX.

The MTXF photometry at $\nu \ge 217\,$GHz of confirmed strongly lensed galaxies in Table\,\ref{tab:confirmed} is presented in Table\,\ref{tab:MTXFconfirmed}. The distributions of the $S_{857}/S_{545}$ and $S_{545}/S_{353}$ flux density ratios of all the 47 confirmed or very likely strongly lensed galaxies are shown in Fig.~\ref{fig:hist_lensed}. All sources but one have $S_{857}/S_{545}< 2.75$.  The highest $S_{857}/S_{545}$ ratio is very uncertain because the source has a low signal-to-noise ratio (SNR) at 545\,GHz ($\hbox{SNR}_{545}\simeq 1.5$): it has an rms error of 74\% and the central value of the ratio is higher than 2.75 by only $0.36\,\sigma$.

The distribution of $S_{545}/S_{353}$ ratios is substantially broader and more uncertain. This is because the SNR is generally low at $353\,$GHz: about one third of sources have $\hbox{SNR}_{353}< 3$ and only 11 have $\hbox{SNR}_{353}> 5$. Thus this ratio does not help much with the selection and we have ignored it. We did not impose any lower limit to the $S_{857}/S_{545}$ ratio because sources with low values  may be at higher redshifts than confirmed strongly lensed galaxies in our sample and thus particularly interesting.

The application of the MTXF technique is quite demanding in terms of computer time. It is therefore not practical to apply it to all sources detected by \textit{Planck} at frequencies $\ge 353\,$GHz. We have therefore chosen to first clean the 353 and 545\,GHz samples as far as possible exploiting the information already available and to obtain the MTXF photometry to cleaned samples only. At 857\,GHz we used the BeeP photometry.

\begin{figure}
\includegraphics[width=\columnwidth]{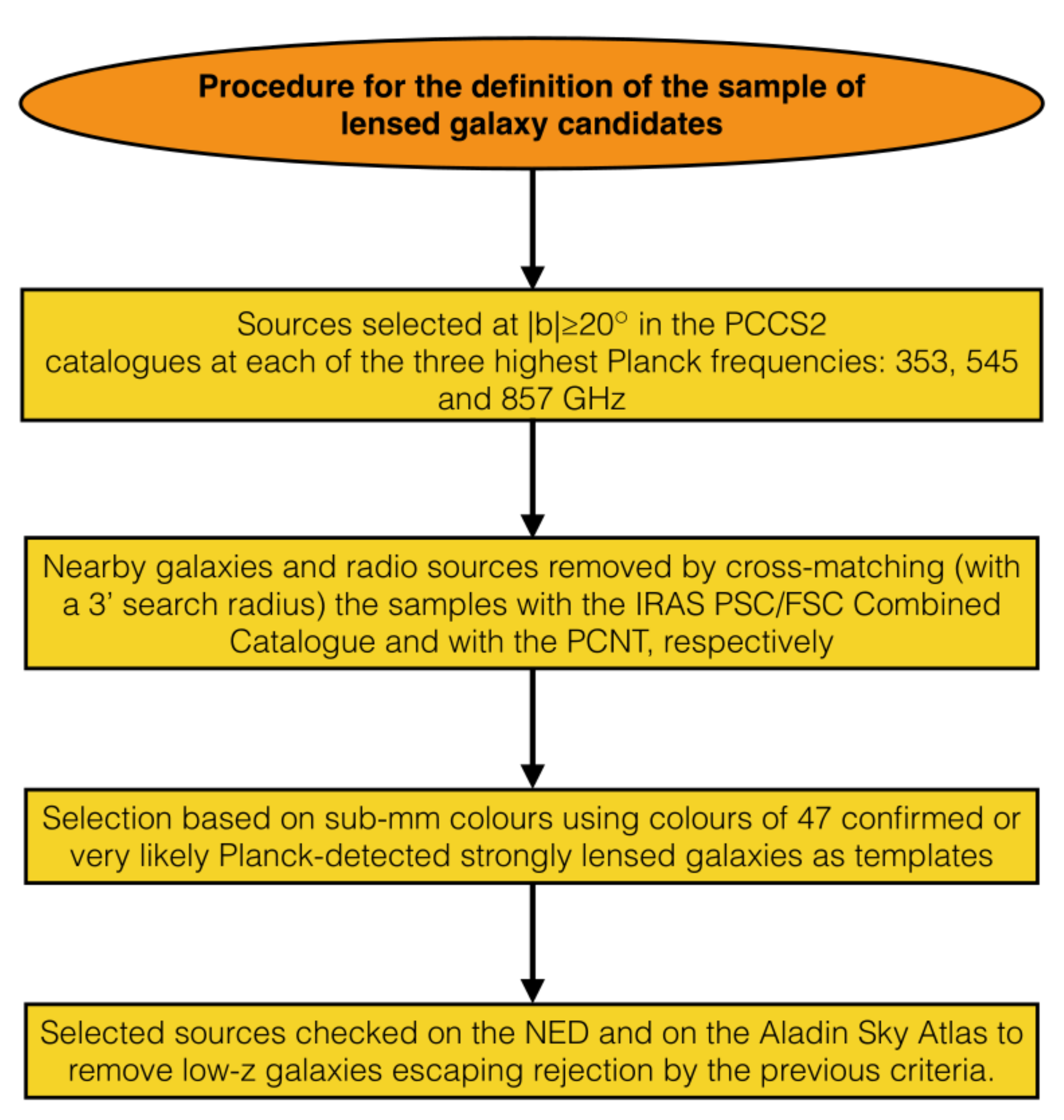}   
\caption{Flowchart summarizing our procedure to select candidate strongly lensed galaxies.}
\label{fig:flowchart}
\end{figure}

Our procedure for selecting \textit{Planck}--detected strongly lensed galaxies is summarized in the flowchart of Fig.~\ref{fig:flowchart}.

\subsection{Selection at 545\,GHz}\label{sect:545}

As mentioned above, we adopted 545\,GHz as our reference frequency. This frequency is expected to maximize the ratio between the number of lensed and unlensed galaxies (although this ratio is nevertheless extremely small) since, compared to the 857\,GHz selection, it favours redder sources, while being more sensitive to dusty galaxies than the 353\,GHz channel.

\begin{table*}
\caption{MTXF photometry of confirmed strongly lensed galaxies listed in Table\,\protect\ref{tab:confirmed}. Flux densities and errors are in mJy, rounded to 1\,mJy. Equatorial coordinates, RA and DEC, J2000,  are in degrees. SN stands for signal-to-noise ratio at the frequency in the subscript.}
\label{tab:MTXFconfirmed}
\begin{adjustbox}{width=\textwidth}
\begin{tabular}{ccccccccccccc}
\hline
  \multicolumn{1}{c}{RA} &
  \multicolumn{1}{c}{DEC} &
  \multicolumn{1}{c}{$S_{217}$}&
  \multicolumn{1}{c}{$S_{353}$}&
  \multicolumn{1}{c}{$S_{545}$}&
  \multicolumn{1}{c}{$S_{857}$}&
  \multicolumn{1}{c}{SN$_{217}$} &
  \multicolumn{1}{c}{SN$_{353}$} &
  \multicolumn{1}{c}{SN$_{545}$} &
  \multicolumn{1}{c}{SN$_{857}$} &
  \multicolumn{1}{c}{$S_{857}/S_{545}$}&
  \multicolumn{1}{c}{$S_{545}/S_{353}$}&
  \multicolumn{1}{c}{$S_{353}/S_{217}$}\\
\hline
 17.4579 & -47.0363 &$ 81 \pm 25$&$248\pm 72$&$ 530\pm105$&$ 854\pm187$ & 3.2 & 3.5 & 5.1 & 4.6 & 1.61 & 2.14 & 3.06\\
 19.1952 & -24.6171 &$ 58 \pm 27$&$264\pm 67$&$ 650\pm111$&$1351\pm260$ & 2.1 & 3.9 & 5.9 & 5.2 & 2.08 & 2.46 & 4.58\\
 21.2802 & -47.3991 &$ 83 \pm 30$&$271\pm 53$&$ 618\pm102$&$1078\pm162$ & 2.8 & 5.1 & 6.1 & 6.7 & 1.74 & 2.28 & 3.26\\
 32.4206 &   0.2624 &$139 \pm 35$&$459\pm 91$&$ 885\pm133$&$1384\pm240$ & 4.0 & 5.0 & 6.6 & 5.8 & 1.56 & 1.93 & 3.31\\
 36.6414 &  23.7581 &$103 \pm 36$&$424\pm100$&$1320\pm163$&$2481\pm413$ & 2.8 & 4.2 & 8.1 & 6.0 & 1.88 & 3.11 & 4.12\\
 46.2946 & -30.6084 &$ 36 \pm 31$&$134\pm 77$&$ 415\pm 98$&$ 694\pm163$ & 1.2 & 1.8 & 4.2 & 4.2 & 1.67 & 3.09 & 3.76\\
117.2154 &  59.6986 &$ 48\pm 36$&$465\pm 70$ & $ 819\pm109$ & $1141\pm250 $ & 1.3 & 6.6 & 7.5 & 4.6 & 1.39 & 1.76 & 9.71\\
131.7091 &  15.0964 &$101\pm 36$&$494\pm 80$ & $1387\pm100$ & $2347\pm190 $ & 2.8 & 6.1 &13.9 &12.3 & 1.69 & 2.81 & 4.89\\
143.0986 &  27.4164 &$143\pm36$&$328\pm 78$ & $ 531\pm 97$ & $ 672\pm188 $ & 4.0 & 4.2 & 5.5 & 3.6 & 1.27 & 1.62 & 2.30\\
163.3439 &  60.8635 &$139\pm31$&$505\pm 64$ & $ 862\pm114$ & $1032\pm215 $ & 4.4 & 7.9 & 7.6 & 4.8 & 1.20 & 1.71 & 3.62\\
163.4709 &   5.9392 &$ 62\pm29$&$370\pm 64$ & $1027\pm108$ & $1852\pm197 $ & 2.1 & 5.7 & 9.5 & 9.4 & 1.80 & 2.77 & 6.02\\
171.8062 &  46.1568 &$ 88\pm27$&$ 81\pm 55$ & $ 142\pm 95$ & $ 529\pm161 $ & 3.3 & 1.5 & 1.5 & 3.3 & 3.73 & 1.75 & 0.92\\
174.5231 &  32.9658 & $44\pm29$&$123\pm 62$ & $ 388\pm112$ & $ 822\pm191 $ & 1.5 & 2.0 & 3.5 & 4.3 & 2.12 & 3.17 & 2.78\\
174.8401 &  20.4146 & $90\pm29$&$222\pm 68$ & $ 505\pm 94$ & $ 585\pm169 $ & 3.1 & 3.3 & 5.3 & 3.5 & 1.16 & 2.27 & 2.46\\
176.6581 &  -0.1923 & $61\pm29$&$318\pm 60$ & $ 601\pm113$ & $1026\pm179 $ & 2.1 & 5.3 & 5.3 & 5.7 & 1.71 & 1.89 & 5.21\\
180.5322 &  53.5779 & $42\pm27$&$168\pm 63$ & $ 544\pm 94$ & $ 909\pm152 $ & 1.6 & 2.7 & 5.8 & 6.0 & 1.67 & 3.24 & 3.96\\
201.6254 &  33.7353 & $47\pm27$&$308\pm 66$ & $ 344\pm101$ & $ 518\pm181 $ & 1.7 & 4.7 & 3.4 & 2.9 & 1.50 & 1.12 & 6.62\\
202.3920 &  22.7242 & $73\pm31$&$247\pm 70$ & $ 886\pm106$ & $1687\pm183 $ & 2.4 & 3.5 & 8.3 & 9.2 & 1.90 & 3.59 & 3.38\\
204.1455 &  49.2205 & $50\pm26$&$240\pm 59$ & $ 537\pm 95$ & $ 946\pm176 $ & 1.9 & 4.1 & 5.7 & 5.4 & 1.76 & 2.24 & 4.79\\
206.1225 &  30.5094 & $40\pm31$&$131\pm 73$ & $ 376\pm 98$ & $ 631\pm169 $ & 1.3 & 1.8 & 3.8 & 3.7 & 1.68 & 2.87 & 3.24\\
222.4941 &  22.6436 & $59\pm37$&$274\pm 97$ & $ 543\pm106$ & $ 912\pm184 $ & 1.6 & 2.8 & 5.1 & 4.9 & 1.68 & 1.98 & 4.66\\
225.6502 &  29.3475 & $72\pm29$&$332\pm 70$ & $ 608\pm 95$ & $ 829\pm169 $ & 2.5 & 4.8 & 6.4 & 4.9 & 1.37 & 1.83 & 4.59\\
236.1350 &  50.3961 & $94\pm27$&$220\pm 49$ & $ 488\pm 92$ & $ 723\pm153 $ & 3.5 & 4.5 & 5.3 & 4.7 & 1.48 & 2.22 & 2.34\\
241.8239 &  73.7844 & $25\pm21$&$132\pm 47$ & $ 459\pm 91$ & $ 951\pm160 $ & 1.2 & 2.8 & 5.0 & 6.0 & 2.07 & 3.48 & 5.25\\
242.3240 &  60.7558 & $29\pm23$&$296\pm 44$ & $ 800\pm 84$ & $1508\pm162 $ & 1.2 & 6.7 & 9.5 & 9.3 & 1.88 & 2.70 & 10.30\\
323.7983 &  -1.0478 & $85\pm34$&$396\pm 78$ & $ 835\pm122$ & $1154\pm224 $ & 2.5 & 5.1 & 6.8 & 5.1 & 1.38 & 2.11 & 4.68\\
353.0980 & -53.9802 & $686 \pm 27$&$290\pm 60$&$ 542\pm120$ & $725\pm 229$ &  2.5 & 4.8 & 4.5 & 3.2 & 1.34 & 1.87 & 4.29\\
\hline\end{tabular}
\end{adjustbox}
\end{table*}

After having removed IRAS galaxies and radio sources we were left with 556 sources that constitute our parent sample.  Applying the $S_{857}/S_{545}< 2.75$ criterion to the MTXF photometry of the parent sample we obtained a sample of 202 candidate strongly lensed. A check made using the NASA/IPAC Extragalactic Database (NED) showed that some of these sources are associated to nearby galaxies that escaped rejection based on the cross-match with the IRAS catalogue. With the help of the Aladin Sky Atlas \citep{Bonnarel2000, BochFernique2014} we picked up 25 such sources that were removed from the sample, leaving 177 candidates, listed in Table\,\ref{tab:MTXFcandidates545GHz}.

We also considered a stricter criterion, $S_{857}/S_{545}< 2.35$. Only 3 confirmed or very likely strongly lensed sources exceed this limit and for two of them the ratio is quite uncertain. We found 116 sources obeying this criterion. The check on the NED showed that 5 of them are local galaxies, leaving 111 sources with $S_{857}/S_{545}< 2.35$ (see Table\,\ref{tab:MTXFcandidates545GHz}).

A test of the efficiency of our selection of strongly lensed candidates was carried out by singling out sources within the SPT area, i.e. $-65^\circ\le \hbox{DEC}\le -40^\circ$ and RA between 20\,h and 7\,h \citep{Everett2020}. In this area there are 15 sources with $S_{857}/S_{545}< 2.75$, 8 of which have an STP counterpart within 3\,arcmin; the latter include 1 confirmed and 4 very likely strongly lensed galaxies. The other 3 SPT matches are classified by \citet{Everett2020} as dusty, unresolved sources and don't have any counterpart in the local universe; therefore they might well be high-$z$ strongly lensed galaxies. In that case our selection efficiency would be $\sim 50\%$. One of the 7 sources lacking an SPT counterpart (at RA=77.30626, DEC= -55.17907) is a PHz source at $z_{\rm phot}=2.61$ \citep{PlanckCollaboration2016highz}; it might be a proto-cluster of dusty galaxies. The other 6 sources might be cirrus, although other possibilities cannot be ruled out.

Among the 177 sources with $S_{857}/S_{545}< 2.75$ there are 13 matches with PHz objects, all with $S_{857}/S_{545}< 1.9$, and 20 matches with GCC's \citep{PlanckCollaboration2016cold_cores}. Seven of the PHz matches and also 7 of the GCC matches are confirmed strongly lensed galaxies (see the last column of Table\,\ref{tab:MTXFcandidates545GHz}).

There are also 7 matches with \textit{Planck} SZ clusters within 5 arcmin\footnote{For cross-matches with cluster catalogues we have adopted a larger search radius ($5'$ instead of $3'$) on account of the larger positional uncertainties of clusters compared to those of non-\textit{Planck} point source catalogues used in this paper.}. Studies of the strong lensing statistics have shown that galaxy clusters contribute substantially to the probability distribution at the very high magnifications typical of \textit{Planck}--detected lensed galaxies \citep{Hilbert2008,  Lima2010, Robertson2020}. In fact, 3 of the confirmed strongly lensed galaxies (at RA, DEC 117.2155, 59.6982; 163.3440, 60.8636; 323.7983, -1.0478) are associated to clusters detected by \textit{Planck} via the Sunyaev-Zeldovich effect \citep{PlanckCollaboration2016SZ}. A cross-match of the full catalogue of PCCS2 detections at 545\,GHz with the \textit{Planck} SZ catalogue didn't yield any other association within $5'$ apart from the chance alignment of the nearby bright galaxy NGC\,4523 which is the obvious identification of the \textit{Planck} source.

We have also cross-matched our candidates with other large catalogues of confirmed galaxy clusters, namely the Massive and Distant Clusters of WISE Survey \citep[MaDCoWS;][]{Gonzalez2019}, the COnstraining Dark Energy with X-ray (CODEX) clusters \citep{Finoguenov2020},  the Meta-Catalogue of X-ray detected Clusters of galaxies \citep[MCXC;][]{Piffaretti2011} and the catalogues of clusters detected via the SZ effect by the SPT \citep{Bleem2015} and by the Atacama Cosmology Telescope \citep[ACT;][]{Hilton2021} surveys. We found 13 matches of our candidate strongly lensed galaxies with clusters in at least one of these catalogues; these source have a ``C'' label in the last column of Table\,\ref{tab:MTXFcandidates545GHz}. A more complete analysis of associations of our lensed candidates with galaxy clusters will be possible when the eROSITA (extended ROentgen Survey with an  Imaging Telescope Array) catalogue, expected to contain $\sim 10^5$ galaxy clusters \citep{Merloni2012} will be available.

Restricting ourselves to $S_{857}/S_{545}< 2.35$, we have 12 sources in the SPT area, with the same 8 matches with SPT sources. This corresponds to a success rate between 5/12 (taking into account only the 5 confirmed or very likely strongly lensed) to 2/3, in the case that all the 8 matches are strongly lensed. The numbers of matches with PHz sources and SZ clusters remain the same, the matches with GCCs decrease to 13.

\subsection{Selection at 857\,GHz}\label{sect:857}

As shown in Table\,\ref{tab:confirmed} some confirmed strongly lensed galaxies were detected by \textit{Planck} only at 857\,GHz. The sample of candidates can therefore be enriched by means of a selection at this frequency which favours lower $z$'s. The approach adopted is analogous to that described above except that we exploited the BeeP photometry which is available for all sources in the PCCS2 857\,GHz list and includes 3000\,GHz ($100\,\mu$m) flux densities extracted from the IRIS map\footnote{The IRIS maps are reprocessed IRAS maps generated by \citet{MivilleDeschenes2005}.} at this frequency. Having photometric data both at higher and at lower frequencies improves the accuracy of the photometry at the selection frequency.

Again we started requiring $|b|\ge 20^\circ$ and removing objects with IRAS or PCNT counterparts, but keeping PCNT sources whose sub-mm emission is dominated by dust. BeeP photometry is available for 23 confirmed or very likely strongly lensed galaxies. Apart from a few outliers, with low signal-to-noise ratios, hence with very uncertain colours, these objects have $S_{857}/S_{545}< 2.75$ (one outlier), $S_{857}/S_{3000}> 2.5$ (two outliers, including the previous one), detection significance $\hbox{SRCSIG}\ge 5$ and $S_{857}<2.65\,$Jy.  We used these limits to select candidate strongly lensed galaxies, except for conservatively relaxing the one on $S_{857}$ to $S_{857}<3\,$Jy and adding the requirement $\hbox{SNRR}>1$ at each frequency, 353, 545 and 857\,GHz\footnote{The SNRR at a given frequency is defined by \citep{PlanckCollaboration2020BEEP} as the source average brightness divided the background standard deviation brightness.  The source detection significance is measured by SRCSIG.}. These criteria yielded a sample of 133 sources, including 21 confirmed or very likely strongly lensed objects (one confirmed strongly lensed galaxy with $S_{857}/S_{545}< 2.75$  has $\hbox{SNR}_{353}<1$).

Dropping the 25 sources included in the 545\,GHz sample (Table~\ref{tab:MTXFcandidates545GHz}) and the 11 local galaxies found checking objects on the NED, we are left with 97 objects, listed in Table~\ref{tab:candidates857GHz}, including 8 confirmed/very likely lensed. The last column of Table~\ref{tab:candidates857GHz} shows that we have 7 matches with the GCC catalogue, one of which is a confirmed strongly lensed galaxy. We also have 2 matches with the PHz catalogue, including a confirmed strongly lensed galaxy. There are no matches with \textit{Planck} SZ clusters. There are however 4 associations within 5 arcmin with at least 1 of the catalogues mentioned in sub-sect.~\ref{sect:545}. These sources are tagged with a ``C'' label in the last column of Table\,\ref{tab:candidates857GHz}.

Eleven of the 97 objects lie in the SPT area, 4 of which have an SPT match including the strongly lensed galaxy at RA=17.4822 deg, DEC=-47.0149. The source at RA=92.9950, DEC=-55.2434 can be identified with the strongly lensed galaxy DES J0611-5514 at $z=0.7$ \citep{Diehl2017}, at an angular separation of 0.76 arcmin, i.e. well within the \textit{Planck} positional error. The other 2 are unresolved by the SPT, don't have any nearby galaxy counterpart and therefore may well be at high $z$, i.e. be strongly lensed. Of the 7 objects without SPT counterpart, three (80.5288, -64.4013; 89.9582, -40.4288; 332.4487, -58.7876) are located in cirrus regions; the other 4 might be CIB fluctuations or proto-clusters of dusty galaxies. Restricting ourselves to $S_{857}/S_{545}< 2.3$ (75 sources) we have 8 sources in the SPT area, including the 4 with SPT counterparts. In this case, the selection efficiency for strongly lensed galaxies is between 25\% and 50\%.

\subsection{Selection at 353\,GHz}\label{sect:353}

The 353\,GHz selection favours higher-$z$ sources. Similarly to what done at 545\,GHz we started from the PCCS2 353\,GHz catalogue selecting objects at $|b|\ge 20^\circ$.  We removed nearby galaxies by dropping matches with the IRAS PSC/FSC Combined Catalogue \citep{Abrahamyan2015} and radio sources by dropping matches with the PCNT \citep{PlanckCollaboration2018PCNT} within a 3\,arcmin search radius, except for sources with sub-mm emission dominated by dust (6 objects). This yielded a sample of 512 sources, including 19 confirmed or very likely strongly lensed galaxies. As expected, the latter have redder colours compared to those selected at the two higher frequencies: all of them have $S_{857}/S_{545}< 2$. For uniformity with previous choices we adopted, to select candidate strongly lensed, $S_{857}/S_{545}< 2.3$.

Removing sources in the 545 and 857\,GHz samples (29 and 5 objects, respectively) we are left with 478 sources, for which we have obtained MTXF photometry. The condition $S_{857}/S_{545}< 2.3$ leaves 228 objects. We further required a SNR at 353 GHz $\hbox{SNR}_{353}>3$ and dropped the 3 sources found, checking on the NED, to be associated with low-$z$ galaxies.  The final sample, containing 104 objects, is presented in Table~\ref{tab:MTXFcandidates353GHz}. Thirteen of the sources in this sample have a \textit{Planck} SZ cluster within 5 arcmin. A cross-match with the catalogues mentioned in sub-sect.~\ref{sect:545} yielded 4 additional associations with galaxy clusters. These 17 sources are tagged with a ``C'' label in the last column of Table\,\ref{tab:MTXFcandidates353GHz}. We also have 6 matches with the PHz catalogue and 7 matches with the GCC catalogue.

Twenty-one of the 104 sources in the final sample lie in the SPT area, 5 of them have SPT counterparts within 3 arcmin. The matches include 2 high-$z$ sources at RA, DEC in degrees (82.2573, -54.6264) and (87.4900,  -53.9362) with $z=3.3689$ and $z= 3.128$, respectively; these can be safely regarded as previously unrecognized \textit{Planck}--detected strongly lensed galaxies. Two sources  (41.3710, -53.0432; 353.0972, -53.9802) have galaxy clusters in the \textit{Planck} SZ catalogue along their lines-of-sight. The \citet{Everett2020} catalogue identifies them with cluster members. However the cluster redshifts, $\simeq 0.3$ and $\simeq 0.4$, respectively, are in the ``zone of avoidance'' of Fig.~\ref{fig:zdistr}, i.e. are either too high or too low to belong to the \textit{Planck} sources. Rather, \textit{Planck} sources are likely background galaxies lensed by the clusters. The fifth source (338.2416, -61.2784) is unresolved by the SPT and hasn't any optical identification, suggesting that it is a high--$z$ dust-enshrouded galaxy.

Two out of the 16 sources in the SPT area lacking an SPT identification (those at 15.7465,  -49.2554; 342.2042,  -44.5310) are associated, in projection,  with \textit{Planck} SZ clusters. A search in the NASA/IPAC Extragalactic Database (NED) has revealed that the first one has, as a possible counterpart, a $z=4.16$ galaxy in the background of the ``El Gordo'' cluster at $z\simeq 0.87$. The region along the line--of--sight of the second is very complex. It contains the cluster Abell\,S1063 at $z=0.3475$ and other overdensities at $z=0.742$, $z\simeq 1.2$ and $z\simeq 3.2$. Several strongly lensed galaxies have been discovered in this region, with $z$ of up to $\simeq 3$. Our source might be one of them or may consist of the summed emission of dusty galaxies in high-$z$ overdensities. The source at (41.3768, -64.3372) can be identified with the PHz source G284-48.60, a candidate high-$z$ proto-cluster of dusty galaxies.

Above $\hbox{SNR}_{353}>4$ we have 41 sources, 14 of which lie in the SPT area, with the same 5 SPT matches and the same associations with SZ clusters.  A selection efficiency between $\sim  35\%$ and $\sim 50\%$ for very likely strongly lensed galaxies is thus indicated in this case.

\section{Discussion and Conclusions}\label{sect:conclusions}

As a result of a systematic search for extreme strongly lensed galaxies in \textit{Planck} catalogues we have produced lists of candidates selected at each of the three highest \textit{Planck} frequencies, 353, 545 and 857\,GHz. Our approach takes advantage of the fact that, without the flux boosting by extreme gravitational lensing, the shallow \textit{Planck} surveys at these frequencies detect only nearby ($z\simlt 0.1$) dusty galaxies. But only  at $z\simgt 1$ there are enough ultraluminous galaxies and a sufficiently large lensing optical depth to allow galaxies with extreme magnifications
\citep[of up to a factor of 50;][]{Canameras2015} to be detectable by \textit{Planck}, taking advantage also of the strongly negative K-correction.  The wide redshift gap between these two populations imply that they have quite different sub-mm colours, high-$z$ galaxies being, on average, substantially redder, although measurement errors blur the difference.

We started by selecting sources at $|b|\ge 20^\circ$ in the PCCS2 catalogues at each of the three frequencies. Next we removed nearby galaxies and radio sources identified by cross-matching the samples with the IRAS PSC/FSC Combined Catalogue \citep{Abrahamyan2015} and with the PCNT \citep{PlanckCollaboration2018PCNT}, respectively, using a 3\,arcmin search radius.

Even with the benefit of extreme gravitational magnifications the flux densities of strongly lensed galaxies are close to the detection limits. Hence they rarely have PCCS2 measurements at all frequencies $\ge 353\,$GHz, as necessary to compute the colours. To deal with this problem we exploited the new BeeP multifrequency photometry \citep{PlanckCollaboration2020BEEP}, which is also constrained by the IRAS 3,000\,GHz photometry, for the sample selected at 857\,GHz.  At the two lower frequencies, at which the BeeP photometry is available only for a subset of sources, we obtained new MTXF multifrequency photometry. The sub-mm colours of 47 confirmed or very likely \textit{Planck}-detected strongly lensed galaxies have been used as a benchmark to select the colour range of lensed candidates. All sources in the colour--selected samples were checked on the NED to remove those associated to low-$z$ galaxies that escaped rejection by the adopted criteria.

Our main sample, selected at 545\,GHz, comprises 177 lensed candidates (Table~\ref{tab:MTXFcandidates545GHz}). The sample selected at 857 GHz contains 97 sources, after having excluded those in the 545\,GHz sample (Table~\ref{tab:candidates857GHz}); the one at 353\,GHz contains 104 sources (Table~\ref{tab:MTXFcandidates353GHz}), excluding those in the other two samples.

A test of the efficiency of our approach in selecting strongly lensed galaxies was made considering that within the area covered by the SPT survey \citep[$\simeq 2,500\,\hbox{deg}^2$;][]{Everett2020}  among sources selected with our method there are from 14 to 19 galaxies which are either classified as strongly lensed by independent data (3) or can be safely regarded as previously unrecognized \textit{Planck}--detected strongly lensed galaxies. Since there are, in total, 47 candidates in that area, the selection efficiency is in the range 30--40\%. We have discussed stricter selection criteria that increase the efficiency to $\simeq 50\%$ at the cost of a somewhat lower completeness: the stricter criteria miss $\simeq 10\%$ of the confirmed plus very likely strongly lensed galaxies recovered by the baseline criteria.

Another test can be made with reference to the H-ATLAS survey covering altogether $\simeq 600\,\hbox{deg}^2$. Thirteen of the objects selected with our method lie within the H-ATLAS fields.
Three of them, those at (RA, DEC: 176.6239, -0.22157; 201.6254, 33.7353; 206.1251, 30.5058) are confirmed strongly lensed galaxies \citep{Negrello2017lensed}, corresponding to an efficiency of $\simeq 23\%$, although with poor statistics ($1\,\sigma$ range 11--46\%).

All the other 10 sources have an H-ATLAS match within 3\,arcmin. The matched sources however have \textit{Herschel}/SPIRE flux densities too faint to be identified with the \textit{Planck} sources (all have $\hbox{F500BEST}< 80\,$mJy). Their colours are red and they don't have possible identifications with low-$z$ galaxies. This suggests that \textit{Planck} detections are high-$z$ overdensities, to which H-ATLAS sources belong.

On the whole, our analysis shows that probably more than 50\% of sources in our samples are not strongly lensed galaxies but a mixture of other objects with ``cold'' spectral energy distributions, such as high-$z$ proto-clusters of dusty galaxies, Galactic cold clumps, CIB fluctuations and Galactic cirrus. Can we exploit specific searches for these objects to clean the samples? As discussed below, unfortunately the answer is no. However, also the discovery of new proto-clusters and GCCs by following up our candidates would be a very interesting scientific result.

\citet{PlanckCollaboration2016highz} published a catalogue of 2151 sources with red sub-mm colours, indicative of $z>2$. These high-$z$ source candidates were extracted from \textit{Planck} high-frequency maps over the 25.8\% of the sky with minimum thermal emission from Galactic dust. The source detection was made using a specific component separation procedure that allowed a much better sensitivity to this class of sources than the PCCS2. \citet{PlanckCollaboration2015clumpsHerschel}, based on the \textit{Herschel} follow-up of 228 \textit{Planck} high-$z$ source candidates, stated that more than 93\% of them are galaxy overdensities, i.e. candidate proto-clusters of dusty galaxies, while 3\% are strongly lensed individual galaxies. At first sight, this suggests that we might exploit the PHz to remove candidate proto-clusters from our samples or, at least, to estimate their fraction.

However, as already pointed out by \citet{PlanckCollaboration2016highz}, the overlap between the PHz and the PCCS2 is extremely small. By cross matching the PHz with the PCCS2 catalogues at 353, 545 and 857\,GHz we found only 29 distinct associations; 21 of them meet our selection criteria, including 8 confirmed (7 sources) or very likely (1 source) strongly lensed galaxies. The fraction of strongly lensed galaxies among PHz sources included in the PCCS2 is thus far larger than in the general PHz catalogue. Taking into account the strong incompleteness of the sample of confirmed strongly lensed galaxies, they may well be the majority in the PCCS2 sub-sample of the PHz, and even more after our selection. Nevertheless, our cross-matches with the SPT and H-ATLAS catalogues have highlighted that some of our high-$z$ candidates may be resolved by the SPT or by \textit{Herschel} and may therefore be proto-clusters not included in the PHz catalogue.

Another population that can meet our selection criteria because of their red sub-mm colours, similar to those of high-$z$ sources, are Galactic cold clumps \citep{PlanckCollaboration2016cold_cores}. Most of them are found close to the Galactic plane, but some were detected also at high Galactic latitude. We found 99 matches within 3 arcmin between the GCC catalogue and the PCCS2. Thirty-four of them, including 8 confirmed strongly lensed galaxies, meet our selection criteria. The substantial fraction of strongly lensed galaxies obviously implies that the presence in the GCC catalogue cannot be a valid criterion for dropping sources from our lists. On the other hand, these objects are of great interest per se since they provide key information on early phases of star formation \citep{PlanckCollaboration2016cold_cores}.

Most published model predictions at sub-mm wavelengths \citep{Perrotta2002, Perrotta2003, Negrello2007, Negrello2017lensed, Bethermin2011, Cai2013} refer to galaxy-galaxy lensing. In this case a maximal magnification $\mu_{\rm max} \simeq 15$ is indicated by the data \citep[see Fig.~7 of][]{Negrello2017lensed}. But the extreme magnifications of \textit{Planck} strongly lensed galaxies can be understood in terms of lensing by galaxy groups or clusters \citep[]{Frye2019}.

\begin{figure}
\includegraphics[width=\columnwidth]{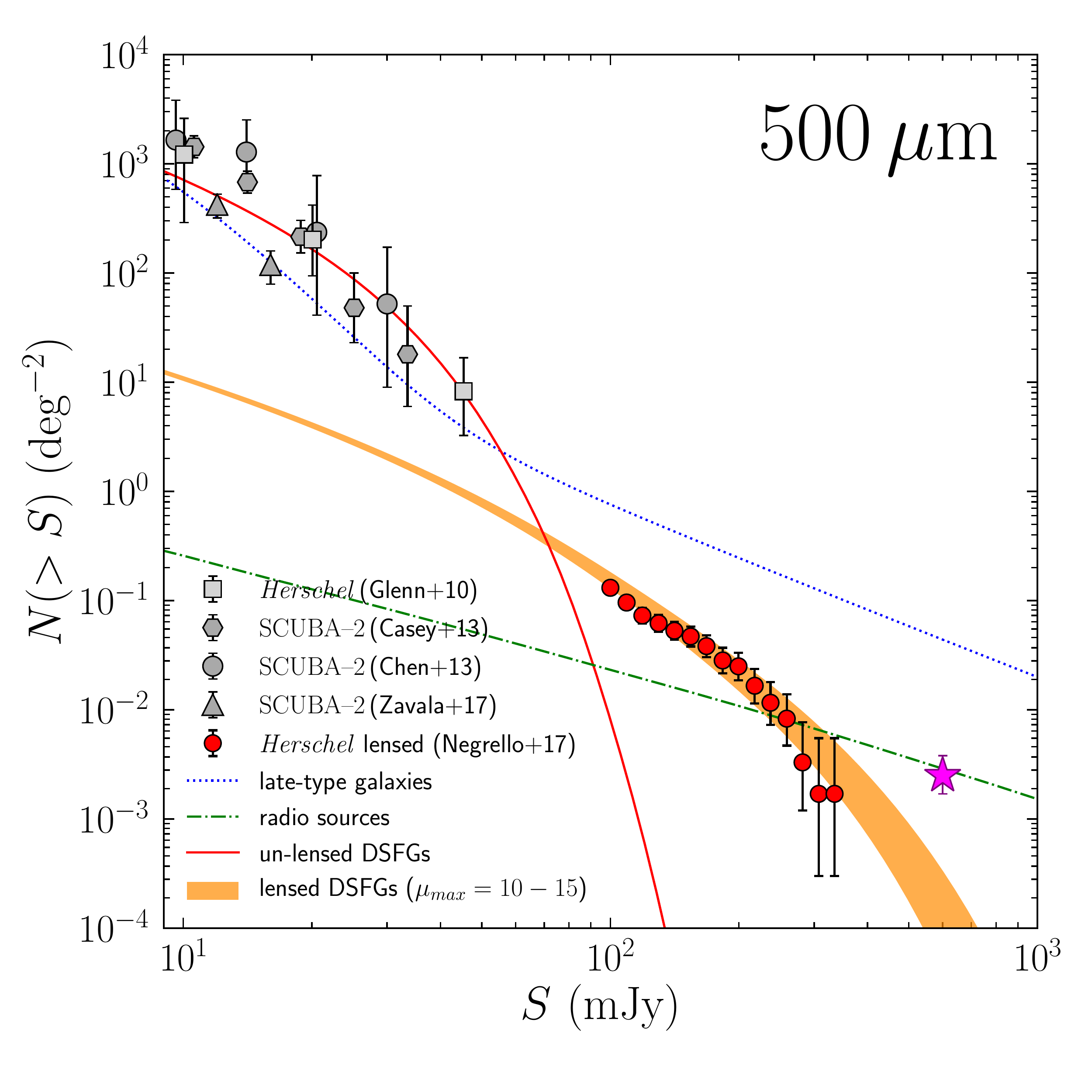}   
\caption{Contributions of different source populations, specified in the legend, to the integral number counts at $500\,\mu$m (600\,GHz), compared with observational data. The counts of late-type, normal and starburst, galaxies and of un-lensed dusty star-forming galaxies (DSFGs), interpreted as proto-spheroidal galaxies in the process of forming the bulk of their stars, are from the \citet{Cai2013} model. The orange band shows the counts of strongly lensed (magnification $\mu > 2$) DSFGs recomputed by \citet{Negrello2017lensed} for magnification cutoffs, $\mu_{\rm max}$, in the range 10--15. The calculations were done exploiting the \citet{Cai2013} model coupled with the \citet{Lapi2012} formalism to deal with galaxy--galaxy lensing.  The counts of radio sources (dot-dashed green line) are from the \citet{Tucci2011} model. The data points are from \citet{Glenn2010}, \citet{Chen2013}, \citet{Casey2013} and \citet{Zavala2017}. The purple star on the bottom-right corner shows our estimate of the counts of \textit{Planck}--detected strongly lensed galaxies. As expected, such counts exceed predictions for galaxy-galaxy lensing, consistent with being mostly contributed by galaxy-cluster lensing. Adapted from Fig.~8 of \citet{Negrello2017lensed}. }
\label{fig:counts}
\end{figure}

Figure~\ref{fig:counts} shows our estimate of the integral number counts of strongly lensed galaxies detected by \textit{Planck}, compared to observed and predicted counts of galaxy-galaxy lensed galaxies at $500\,\mu$m (600\,GHz). This estimate was obtained as follows. The 90\% completeness flux density limit at our reference frequency of 545\,GHz is $S_{545,\rm lim}\simeq 500\,$mJy \citep{PCCS2, Maddox2018}. We can identify with good completeness and reliability \textit{Planck}--detected strongly lensed galaxies only in areas covered by deeper, higher angular resolution surveys, i.e. over the $3,100\,\hbox{deg}^2$ surveyed by the SPT plus H-ATLAS. We have 11 such sources in this area, 7 of which have a flux density above $S_{545,\rm lim}$. For a typical redshift $z\simeq 2$ the proto-spheroidal SED by \citet{Cai2013} yields $S_{500\,\mu\rm m}/S_{545\,\rm GHz} = 1.2$. After correcting for the 10\% incompleteness we find $N(>500\,\rm mJy) =  2.5(+1.3,-0.8)\times 10^{-3}\,\hbox{deg}^{-2}$ at $500\,\mu$m, with Poisson errors computed following \citet{Gehrels1986}.

The figure shows that indeed the counts of \textit{Planck} strongly lensed galaxies exceed the extrapolation of observed counts and model predictions, both referring to galaxy-galaxy lensing. The important role of galaxy-cluster lensing is supported by the results of our cross-match with cluster catalogues. We have found a total of 20 associations of our lensed candidate samples with the \textit{Planck} SZ catalogue \citep{PlanckCollaboration2016SZ}. Such catalogue is not deep enough for a thorough search for associations, but it the best one available at the moment, while waiting for the eROSITA cluster survey; deeper catalogues cover a small fraction of the sky. The \textit{Planck} SZ catalogue contains 1653 clusters spread over $\simeq 34,500\,\hbox{deg}^2$. Their mean surface density is thus of $\simeq 0.048\,\hbox{clusters}\,\hbox{deg}^{-2}$. The number of chance associations within the total searched area $\simeq \pi(5/60)^2 \times N_{\rm cand}\simeq 8.2\,\hbox{deg}^2$, $N_{\rm cand}=377$ being the sum of candidates in our 3 samples, is 0.39. The Poisson probability of having by chance 20 associations when the expected number is 0.39 is vanishingly small, confirming that our candidates have a strong excess of foreground clusters. 

Unfortunately, calculations including magnifications by galaxy clusters \citep{Lima2010, Er2013} use outdated models and do not extend their predictions to the flux densities of interest here. Hence, a proper comparison of our count estimate with model predictions cannot be done at this stage.

Summing up, how many \textit{Planck}-detected strongly lensed galaxies can we expect? Some useful indications are provided by the cross-matches with the SPT and H-ATLAS surveys. The two surveys contain a total of 17--22 confirmed or very likely strongly lensed galaxies detected by \textit{Planck} over an area of $3100\,\hbox{deg}^2$. This corresponds to 149--192 such objects over the full area at $|b|>20^\circ$ ($\simeq 2.71\times 10^4\,\hbox{deg}^2$).

We conclude that \textit{Planck}--detected strongly lensed galaxies constitute a rich enough sample to provide, via high resolution spectro-photometric follow up, a uniquely detailed view of the internal structure and kinematics of galaxies across the peak of the cosmic star formation. Moreover, this sample can be exploited to study the spatial distribution of dark and luminous mass in galaxies or galaxy clusters acting as lenses over a broader redshift range than is possible with optically selected strongly lensed galaxies.

Follow-up work on candidates has started. SCUBA-2 photometric observations at $850\,\mu$m (353\,GHz) of a preliminary selection (Proposal ID: M19BP010, P.I.: M. Negrello) have been already mentioned; 4 of the detected sources were observed spectroscopically with the Northern Extended Millimeter Array (NOEMA; Proposal ID: S20BQ, P.I.: M. Negrello). Other samples were observed with the Australia Telescope Compact Array (ATCA) at 5.5 and 94\,GHz (project ID\,C3301, P.I.: M. Bonato, 95\,h of observing time) and with IRAM's second generation Neel-IRAM-KID-Array (NIKA\,2) at 1 and 2\,mm (project ID\,212-19, P.I.: M. Bonato, 13.5\,h of observing time). The analysis of these data is in progress.

\begin{acknowledgements}
We are grateful to the anonymous referee for useful comments.
DH thanks the Spanish Ministerio de Ciencia, Innovaci\'on y Universidades for partial financial support under project PGC2018-101814-B-I00.
MB acknowledges support from INAF under PRIN SKA/CTA FORECaST, from the Ministero degli Affari Esteri e della Cooperazione Internazionale - Direzione
Generale per la Promozione del Sistema Paese Progetto di Grande Rilevanza ZA18GR02 and from the South African Department of Science and Technology's
National Research Foundation (DST-NRF Grant Number 113121).
This research has made use of the \textit{Planck} Legacy Archive (\url{https://www.cosmos.esa.int/web/planck/pla}), of the TOPCAT software \citep{TOPCAT2005}, of the Aladin Sky Atlas developed at CDS, Strasbourg Observatory, France, and of the NASA/IPAC Extragalactic Database (NED). \textit{Planck} is an ESA science mission with instruments and contributions directly funded by ESA Member States, NASA, and Canada. The NED is operated by the Jet Propulsion Laboratory, California Institute of Technology, under contract with the National Aeronautics and Space Administration.
\end{acknowledgements}

%
%


\begin{thebibliography}{101}
\expandafter\ifx\csname natexlab\endcsname\relax\def\natexlab#1{#1}\fi

\bibitem[{{Abrahamyan} {et~al.}(2015){Abrahamyan}, {Mickaelian}, \&
  {Knyazyan}}]{Abrahamyan2015}
{Abrahamyan}, H.~V., {Mickaelian}, A.~M., \& {Knyazyan}, A.~V. 2015, Astronomy
  and Computing, 10, 99

\bibitem[{{Bakx} {et~al.}(2020){Bakx}, {Eales}, \& {Amvrosiadis}}]{Bakx2020}
{Bakx}, T. J.~L.~C., {Eales}, S., \& {Amvrosiadis}, A. 2020, \mnras, 493, 4276

\bibitem[{{B{\'e}thermin} {et~al.}(2011){B{\'e}thermin}, {Dole}, {Lagache}, {Le
  Borgne}, \& {Penin}}]{Bethermin2011}
{B{\'e}thermin}, M., {Dole}, H., {Lagache}, G., {Le Borgne}, D., \& {Penin}, A.
  2011, \aap, 529, A4

\bibitem[{{Bleem} {et~al.}(2015){Bleem}, {Stalder}, {de Haan}, {Aird}, {Allen},
  {Applegate}, {Ashby}, {Bautz}, {Bayliss}, {Benson}, {Bocquet}, {Brodwin},
  {Carlstrom}, {Chang}, {Chiu}, {Cho}, {Clocchiatti}, {Crawford}, {Crites},
  {Desai}, {Dietrich}, {Dobbs}, {Foley}, {Forman}, {George}, {Gladders},
  {Gonzalez}, {Halverson}, {Hennig}, {Hoekstra}, {Holder}, {Holzapfel},
  {Hrubes}, {Jones}, {Keisler}, {Knox}, {Lee}, {Leitch}, {Liu}, {Lueker},
  {Luong-Van}, {Mantz}, {Marrone}, {McDonald}, {McMahon}, {Meyer}, {Mocanu},
  {Mohr}, {Murray}, {Padin}, {Pryke}, {Reichardt}, {Rest}, {Ruel}, {Ruhl},
  {Saliwanchik}, {Saro}, {Sayre}, {Schaffer}, {Schrabback}, {Shirokoff},
  {Song}, {Spieler}, {Stanford}, {Staniszewski}, {Stark}, {Story}, {Stubbs},
  {Vanderlinde}, {Vieira}, {Vikhlinin}, {Williamson}, {Zahn}, \&
  {Zenteno}}]{Bleem2015}
{Bleem}, L.~E., {Stalder}, B., {de Haan}, T., {et~al.} 2015, \apjs, 216, 27

\bibitem[{{Boch} \& {Fernique}(2014)}]{BochFernique2014}
{Boch}, T. \& {Fernique}, P. 2014, in Astronomical Society of the Pacific
  Conference Series, Vol. 485, Astronomical Data Analysis Software and Systems
  XXIII, ed. N.~{Manset} \& P.~{Forshay}, 277

\bibitem[{{Bonnarel} {et~al.}(2000){Bonnarel}, {Fernique}, {Bienaym{\'e}},
  {Egret}, {Genova}, {Louys}, {Ochsenbein}, {Wenger}, \&
  {Bartlett}}]{Bonnarel2000}
{Bonnarel}, F., {Fernique}, P., {Bienaym{\'e}}, O., {et~al.} 2000, \aaps, 143,
  33

\bibitem[{{Bourne} {et~al.}(2016){Bourne}, {Dunne}, {Maddox}, {Dye},
  {Furlanetto}, {Hoyos}, {Smith}, {Eales}, {Smith}, {Valiante}, {Alpaslan},
  {Andrae}, {Baldry}, {Cluver}, {Cooray}, {Driver}, {Dunlop}, {Grootes},
  {Ivison}, {Jarrett}, {Liske}, {Madore}, {Popescu}, {Robotham}, {Rowlands},
  {Seibert}, {Thompson}, {Tuffs}, {Viaene}, \& {Wright}}]{Bourne2016}
{Bourne}, N., {Dunne}, L., {Maddox}, S.~J., {et~al.} 2016, \mnras, 462, 1714

\bibitem[{{Ca{\~n}ameras} {et~al.}(2017{\natexlab{a}}){Ca{\~n}ameras},
  {Nesvadba}, {Kneissl}, {Frye}, {Gavazzi}, {Koenig}, {Le Floc'h}, {Limousin},
  {Oteo}, \& {Scott}}]{Canameras2017ALMA}
{Ca{\~n}ameras}, R., {Nesvadba}, N., {Kneissl}, R., {et~al.}
  2017{\natexlab{a}}, \aap, 604, A117

\bibitem[{{Ca{\~n}ameras} {et~al.}(2015){Ca{\~n}ameras}, {Nesvadba}, {Guery},
  {McKenzie}, {K{\"o}nig}, {Petitpas}, {Dole}, {Frye}, {Flores-Cacho},
  {Montier}, {Negrello}, {Beelen}, {Boone}, {Dicken}, {Lagache}, {Le Floc'h},
  {Altieri}, {B{\'e}thermin}, {Chary}, {de Zotti}, {Giard}, {Kneissl}, {Krips},
  {Malhotra}, {Martinache}, {Omont}, {Pointecouteau}, {Puget}, {Scott},
  {Soucail}, {Valtchanov}, {Welikala}, \& {Yan}}]{Canameras2015}
{Ca{\~n}ameras}, R., {Nesvadba}, N.~P.~H., {Guery}, D., {et~al.} 2015, \aap,
  581, A105

\bibitem[{{Ca{\~n}ameras} {et~al.}(2021){Ca{\~n}ameras}, {Nesvadba}, {Kneissl},
  {K{\"o}nig}, {Yang}, {Beelen}, {Hill}, {Le Floc'h}, \&
  {Scott}}]{Canameras2020}
{Ca{\~n}ameras}, R., {Nesvadba}, N.~P.~H., {Kneissl}, R., {et~al.} 2021, \aap,
  645, A45

\bibitem[{{Ca{\~n}ameras} {et~al.}(2017{\natexlab{b}}){Ca{\~n}ameras},
  {Nesvadba}, {Kneissl}, {Limousin}, {Gavazzi}, {Scott}, {Dole}, {Frye},
  {Koenig}, {Le Floc'h}, \& {Oteo}}]{Canameras2017}
{Ca{\~n}ameras}, R., {Nesvadba}, N.~P.~H., {Kneissl}, R., {et~al.}
  2017{\natexlab{b}}, \aap, 600, L3

\bibitem[{{Ca{\~n}ameras} {et~al.}(2018{\natexlab{a}}){Ca{\~n}ameras},
  {Nesvadba}, {Limousin}, {Dole}, {Kneissl}, {Koenig}, {Le Floc'h}, {Petitpas},
  \& {Scott}}]{Canameras2018a}
{Ca{\~n}ameras}, R., {Nesvadba}, N.~P.~H., {Limousin}, M., {et~al.}
  2018{\natexlab{a}}, \aap, 620, A60

\bibitem[{{Ca{\~n}ameras} {et~al.}(2018{\natexlab{b}}){Ca{\~n}ameras}, {Yang},
  {Nesvadba}, {Beelen}, {Kneissl}, {Koenig}, {Le Floc'h}, {Limousin},
  {Malhotra}, {Omont}, \& {Scott}}]{Canameras2018b}
{Ca{\~n}ameras}, R., {Yang}, C., {Nesvadba}, N.~P.~H., {et~al.}
  2018{\natexlab{b}}, \aap, 620, A61

\bibitem[{{Cai} {et~al.}(2013){Cai}, {Lapi}, {Xia}, {De Zotti}, {Negrello},
  {Gruppioni}, {Rigby}, {Castex}, {Delabrouille}, \& {Danese}}]{Cai2013}
{Cai}, Z.-Y., {Lapi}, A., {Xia}, J.-Q., {et~al.} 2013, \apj, 768, 21

\bibitem[{{Cao} {et~al.}(2020){Cao}, {Li}, {Shu}, {Mao}, {Kneib}, \&
  {Gao}}]{Cao2020}
{Cao}, X., {Li}, R., {Shu}, Y., {et~al.} 2020, \mnras, 499, 3610

\bibitem[{{Casey} {et~al.}(2013){Casey}, {Chen}, {Cowie}, {Barger}, {Capak},
  {Ilbert}, {Koss}, {Lee}, {Le Floc'h}, {Sanders}, \& {Williams}}]{Casey2013}
{Casey}, C.~M., {Chen}, C.-C., {Cowie}, L.~L., {et~al.} 2013, \mnras, 436, 1919

\bibitem[{{Chen} {et~al.}(2013){Chen}, {Cowie}, {Barger}, {Casey}, {Lee},
  {Sanders}, {Wang}, \& {Williams}}]{Chen2013}
{Chen}, C.-C., {Cowie}, L.~L., {Barger}, A.~J., {et~al.} 2013, \apj, 776, 131

\bibitem[{{Dannerbauer} {et~al.}(2019){Dannerbauer}, {Harrington},
  {D{\'\i}az-S{\'a}nchez}, {Iglesias-Groth}, {Rebolo}, {Genova-Santos}, \&
  {Krips}}]{Dannerbauer2019}
{Dannerbauer}, H., {Harrington}, K., {D{\'\i}az-S{\'a}nchez}, A., {et~al.}
  2019, \aj, 158, 34

\bibitem[{{Delabrouille} {et~al.}(2013){Delabrouille}, {Betoule}, {Melin},
  {Miville-Desch{\^e}nes}, {Gonzalez-Nuevo}, {Le Jeune}, {Castex}, {de Zotti},
  {Basak}, {Ashdown}, {Aumont}, {Baccigalupi}, {Banday}, {Bernard}, {Bouchet},
  {Clements}, {da Silva}, {Dickinson}, {Dodu}, {Dolag}, {Elsner}, {Fauvet},
  {Fa{\"y}}, {Giardino}, {Leach}, {Lesgourgues}, {Liguori},
  {Mac{\'\i}as-P{\'e}rez}, {Massardi}, {Matarrese}, {Mazzotta}, {Montier},
  {Mottet}, {Paladini}, {Partridge}, {Piffaretti}, {Prezeau}, {Prunet},
  {Ricciardi}, {Roman}, {Schaefer}, \& {Toffolatti}}]{Delabrouille2013}
{Delabrouille}, J., {Betoule}, M., {Melin}, J.~B., {et~al.} 2013, \aap, 553,
  A96

\bibitem[{{D{\'{\i}}az-S{\'a}nchez} {et~al.}(2017){D{\'{\i}}az-S{\'a}nchez},
  {Iglesias-Groth}, {Rebolo}, \& {Dannerbauer}}]{DiazSanchez2017}
{D{\'{\i}}az-S{\'a}nchez}, A., {Iglesias-Groth}, S., {Rebolo}, R., \&
  {Dannerbauer}, H. 2017, \apjl, 843, L22

\bibitem[{{Diehl} {et~al.}(2017){Diehl}, {Buckley-Geer}, {Lindgren}, {Nord},
  {Gaitsch}, {Gaitsch}, {Lin}, {Allam}, {Collett}, {Furlanetto}, {Gill},
  {More}, {Nightingale}, {Odden}, {Pellico}, {Tucker}, {da Costa}, {Fausti
  Neto}, {Kuropatkin}, {Soares-Santos}, {Welch}, {Zhang}, {Frieman}, {Abdalla},
  {Annis}, {Benoit-L{\'e}vy}, {Bertin}, {Brooks}, {Burke}, {Carnero Rosell},
  {Carrasco Kind}, {Carretero}, {Cunha}, {D'Andrea}, {Desai}, {Dietrich},
  {Drlica-Wagner}, {Evrard}, {Finley}, {Flaugher}, {Garc{\'\i}a-Bellido},
  {Gerdes}, {Goldstein}, {Gruen}, {Gruendl}, {Gschwend}, {Gutierrez}, {James},
  {Kuehn}, {Kuhlmann}, {Lahav}, {Li}, {Lima}, {Maia}, {Marshall}, {Menanteau},
  {Miquel}, {Nichol}, {Nugent}, {Ogando}, {Plazas}, {Reil}, {Romer}, {Sako},
  {Sanchez}, {Santiago}, {Scarpine}, {Schindler}, {Schubnell},
  {Sevilla-Noarbe}, {Sheldon}, {Smith}, {Sobreira}, {Suchyta}, {Swanson},
  {Tarle}, {Thomas}, {Walker}, \& {DES Collaboration}}]{Diehl2017}
{Diehl}, H.~T., {Buckley-Geer}, E.~J., {Lindgren}, K.~A., {et~al.} 2017, \apjs,
  232, 15

\bibitem[{{Eales} {et~al.}(2010){Eales}, {Dunne}, {Clements}, {Cooray}, {De
  Zotti}, {Dye}, {Ivison}, {Jarvis}, {Lagache}, {Maddox}, {Negrello},
  {Serjeant}, {Thompson}, {Van Kampen}, {Amblard}, {Andreani}, {Baes},
  {Beelen}, {Bendo}, {Benford}, {Bertoldi}, {Bock}, {Bonfield}, {Boselli},
  {Bridge}, {Buat}, {Burgarella}, {Carlberg}, {Cava}, {Chanial}, {Charlot},
  {Christopher}, {Coles}, {Cortese}, {Dariush}, {da Cunha}, {Dalton}, {Danese},
  {Dannerbauer}, {Driver}, {Dunlop}, {Fan}, {Farrah}, {Frayer}, {Frenk},
  {Geach}, {Gardner}, {Gomez}, {Gonz{\'a}lez-Nuevo}, {Gonz{\'a}lez-Solares},
  {Griffin}, {Hardcastle}, {Hatziminaoglou}, {Herranz}, {Hughes}, {Ibar},
  {Jeong}, {Lacey}, {Lapi}, {Lawrence}, {Lee}, {Leeuw}, {Liske},
  {L{\'o}pez-Caniego}, {M{\"u}ller}, {Nandra}, {Panuzzo}, {Papageorgiou},
  {Patanchon}, {Peacock}, {Pearson}, {Phillipps}, {Pohlen}, {Popescu},
  {Rawlings}, {Rigby}, {Rigopoulou}, {Robotham}, {Rodighiero}, {Sansom},
  {Schulz}, {Scott}, {Smith}, {Sibthorpe}, {Smail}, {Stevens}, {Sutherland},
  {Takeuchi}, {Tedds}, {Temi}, {Tuffs}, {Trichas}, {Vaccari}, {Valtchanov},
  {van der Werf}, {Verma}, {Vieria}, {Vlahakis}, \& {White}}]{Eales2010}
{Eales}, S., {Dunne}, L., {Clements}, D., {et~al.} 2010, \pasp, 122, 499

\bibitem[{{Enia} {et~al.}(2018){Enia}, {Negrello}, {Gurwell}, {Dye},
  {Rodighiero}, {Massardi}, {De Zotti}, {Franceschini}, {Cooray}, {van der
  Werf}, {Birkinshaw}, {Micha{\l}owski}, \& {Oteo}}]{Enia2018}
{Enia}, A., {Negrello}, M., {Gurwell}, M., {et~al.} 2018, \mnras, 475, 3467

\bibitem[{{Er} {et~al.}(2013){Er}, {Li}, {Mao}, \& {Cao}}]{Er2013}
{Er}, X., {Li}, G., {Mao}, S., \& {Cao}, L. 2013, \mnras, 430, 1423

\bibitem[{{Everett} {et~al.}(2020){Everett}, {Zhang}, {Crawford}, {Vieira},
  {Aravena}, {Archipley}, {Austermann}, {Benson}, {Bleem}, {Carlstrom},
  {Chang}, {Chapman}, {Crites}, {de Haan}, {Dobbs}, {George}, {Halverson},
  {Harrington}, {Holder}, {Holzapfel}, {Hrubes}, {Knox}, {Lee}, {Luong-Van},
  {Mangian}, {Marrone}, {McMahon}, {Meyer}, {Mocanu}, {Mohr}, {Natoli},
  {Padin}, {Pryke}, {Reichardt}, {Reuter}, {Ruhl}, {Sayre}, {Schaffer},
  {Shirokoff}, {Spilker}, {Stalder}, {Staniszewski}, {Stark}, {Story},
  {Switzer}, {Vanderlinde}, {Wei{\ss}}, \& {Williamson}}]{Everett2020}
{Everett}, W.~B., {Zhang}, L., {Crawford}, T.~M., {et~al.} 2020, \apj, 900, 55

\bibitem[{{Finoguenov} {et~al.}(2020){Finoguenov}, {Rykoff}, {Clerc},
  {Costanzi}, {Hagstotz}, {Ider Chitham}, {Kiiveri}, {Kirkpatrick}, {Capasso},
  {Comparat}, {Damsted}, {Dupke}, {Erfanianfar}, {Patrick Henry}, {Kaefer},
  {Kneib}, {Lindholm}, {Rozo}, {van Waerbeke}, \& {Weller}}]{Finoguenov2020}
{Finoguenov}, A., {Rykoff}, E., {Clerc}, N., {et~al.} 2020, \aap, 638, A114

\bibitem[{{Frye} {et~al.}(2019){Frye}, {Pascale}, {Qin}, {Zitrin}, {Diego},
  {Walth}, {Yan}, {Conselice}, {Alpaslan}, {Bauer}, {Busoni}, {Coe}, {Cohen},
  {Dole}, {Donahue}, {Georgiev}, {Jansen}, {Limousin}, {Livermore}, {Norman},
  {Rabien}, \& {Windhorst}}]{Frye2019}
{Frye}, B.~L., {Pascale}, M., {Qin}, Y., {et~al.} 2019, \apj, 871, 51

\bibitem[{{Fu} {et~al.}(2012){Fu}, {Jullo}, {Cooray}, {Bussmann}, {Ivison},
  {P{\'e}rez-Fournon}, {Djorgovski}, {Scoville}, {Yan}, {Riechers}, {Aguirre},
  {Auld}, {Baes}, {Baker}, {Bradford}, {Cava}, {Clements}, {Dannerbauer},
  {Dariush}, {De Zotti}, {Dole}, {Dunne}, {Dye}, {Eales}, {Frayer}, {Gavazzi},
  {Gurwell}, {Harris}, {Herranz}, {Hopwood}, {Hoyos}, {Ibar}, {Jarvis}, {Kim},
  {Leeuw}, {Lupu}, {Maddox}, {Mart{\'{\i}}nez-Navajas}, {Micha{\l}owski},
  {Negrello}, {Omont}, {Rosenman}, {Scott}, {Serjeant}, {Smail}, {Swinbank},
  {Valiante}, {Verma}, {Vieira}, {Wardlow}, \& {van der Werf}}]{Fu2012}
{Fu}, H., {Jullo}, E., {Cooray}, A., {et~al.} 2012, \apj, 753, 134

\bibitem[{{Fujimoto} {et~al.}(2018){Fujimoto}, {Ouchi}, {Kohno}, {Yamaguchi},
  {Hatsukade}, {Ueda}, {Shibuya}, {Inoue}, {Oogi}, {Toft},
  {G{\'o}mez-Guijarro}, {Wang}, {Espada}, {Nagao}, {Tanaka}, {Ao}, {Umehata},
  {Taniguchi}, {Nakanishi}, {Rujopakarn}, {Ivison}, {Wang}, {Lee}, {Tadaki},
  {Tamura}, \& {Dunlop}}]{Fujimoto2018}
{Fujimoto}, S., {Ouchi}, M., {Kohno}, K., {et~al.} 2018, \apj, 861, 7

\bibitem[{{Fujimoto} {et~al.}(2020){Fujimoto}, {Silverman}, {Bethermin},
  {Ginolfi}, {Jones}, {Le F{\`e}vre}, {Dessauges-Zavadsky}, {Rujopakarn},
  {Faisst}, {Fudamoto}, {Cassata}, {Morselli}, {Maiolino}, {Schaerer}, {Capak},
  {Yan}, {Vallini}, {Toft}, {Loiacono}, {Zamorani}, {Talia}, {Narayanan},
  {Hathi}, {Lemaux}, {Boquien}, {Amorin}, {Ibar}, {Koekemoer},
  {M{\'e}ndez-Hern{\'a}ndez}, {Bardelli}, {Vergani}, {Zucca}, {Romano}, \&
  {Cimatti}}]{Fujimoto2020}
{Fujimoto}, S., {Silverman}, J.~D., {Bethermin}, M., {et~al.} 2020, \apj, 900,
  1

\bibitem[{{Gehrels}(1986)}]{Gehrels1986}
{Gehrels}, N. 1986, \apj, 303, 336

\bibitem[{{Glenn} {et~al.}(2010){Glenn}, {Conley}, {B{\'e}thermin}, {Altieri},
  {Amblard}, {Arumugam}, {Aussel}, {Babbedge}, {Blain}, {Bock}, {Boselli},
  {Buat}, {Castro-Rodr{\'\i}guez}, {Cava}, {Chanial}, {Clements}, {Conversi},
  {Cooray}, {Dowell}, {Dwek}, {Eales}, {Elbaz}, {Ellsworth-Bowers}, {Fox},
  {Franceschini}, {Gear}, {Griffin}, {Halpern}, {Hatziminaoglou}, {Ibar},
  {Isaak}, {Ivison}, {Lagache}, {Laurent}, {Levenson}, {Lu}, {Madden},
  {Maffei}, {Mainetti}, {Marchetti}, {Marsden}, {Nguyen}, {O'Halloran},
  {Oliver}, {Omont}, {Page}, {Panuzzo}, {Papageorgiou}, {Pearson},
  {P{\'e}rez-Fournon}, {Pohlen}, {Rigopoulou}, {Rizzo}, {Roseboom},
  {Rowan-Robinson}, {Portal}, {Schulz}, {Scott}, {Seymour}, {Shupe}, {Smith},
  {Stevens}, {Symeonidis}, {Trichas}, {Tugwell}, {Vaccari}, {Valtchanov},
  {Vieira}, {Vigroux}, {Wang}, {Ward}, {Wright}, {Xu}, \& {Zemcov}}]{Glenn2010}
{Glenn}, J., {Conley}, A., {B{\'e}thermin}, M., {et~al.} 2010, \mnras, 409, 109

\bibitem[{{Gonzalez} {et~al.}(2019){Gonzalez}, {Gettings}, {Brodwin},
  {Eisenhardt}, {Stanford}, {Wylezalek}, {Decker}, {Marrone}, {Moravec},
  {O'Donnell}, {Stalder}, {Stern}, {Abdulla}, {Brown}, {Carlstrom}, {Chambers},
  {Hayden}, {Lin}, {Magnier}, {Masci}, {Mantz}, {McDonald}, {Mo}, {Perlmutter},
  {Wright}, \& {Zeimann}}]{Gonzalez2019}
{Gonzalez}, A.~H., {Gettings}, D.~P., {Brodwin}, M., {et~al.} 2019, \apjs, 240,
  33

\bibitem[{{Gonz{\'a}lez-Nuevo} {et~al.}(2012){Gonz{\'a}lez-Nuevo}, {Lapi},
  {Fleuren}, {Bressan}, {Danese}, {De Zotti}, {Negrello}, {Cai}, {Fan},
  {Sutherland}, {Baes}, {Baker}, {Clements}, {Cooray}, {Dannerbauer}, {Dunne},
  {Dye}, {Eales}, {Frayer}, {Harris}, {Ivison}, {Jarvis}, {Micha{\l}owski},
  {L{\'o}pez-Caniego}, {Rodighiero}, {Rowlands}, {Serjeant}, {Scott}, {van der
  Werf}, {Auld}, {Buttiglione}, {Cava}, {Dariush}, {Fritz}, {Hopwood}, {Ibar},
  {Maddox}, {Pascale}, {Pohlen}, {Rigby}, {Smith}, \&
  {Temi}}]{GonzalezNuevo2012}
{Gonz{\'a}lez-Nuevo}, J., {Lapi}, A., {Fleuren}, S., {et~al.} 2012, \apj, 749,
  65

\bibitem[{{Gonz{\'a}lez-Nuevo} {et~al.}(2019){Gonz{\'a}lez-Nuevo}, {Su{\'a}rez
  G{\'o}mez}, {Bonavera}, {S{\'a}nchez-Lasheras}, {Arg{\"u}eso}, {Toffolatti},
  {Herranz}, {Gonz{\'a}lez-Guti{\'e}rrez}, {Garc{\'\i}a Riesgo}, \& {de Cos
  Juez}}]{GonzalezNuevo2019}
{Gonz{\'a}lez-Nuevo}, J., {Su{\'a}rez G{\'o}mez}, S.~L., {Bonavera}, L.,
  {et~al.} 2019, \aap, 627, A31

\bibitem[{{Gruppioni} {et~al.}(2013){Gruppioni}, {Pozzi}, {Rodighiero},
  {Delvecchio}, {Berta}, {Pozzetti}, {Zamorani}, {Andreani}, {Cimatti},
  {Ilbert}, {Le Floc'h}, {Lutz}, {Magnelli}, {Marchetti}, {Monaco}, {Nordon},
  {Oliver}, {Popesso}, {Riguccini}, {Roseboom}, {Rosario}, {Sargent},
  {Vaccari}, {Altieri}, {Aussel}, {Bongiovanni}, {Cepa}, {Daddi},
  {Dom{\'\i}nguez-S{\'a}nchez}, {Elbaz}, {F{\"o}rster Schreiber}, {Genzel},
  {Iribarrem}, {Magliocchetti}, {Maiolino}, {Poglitsch}, {P{\'e}rez
  Garc{\'\i}a}, {Sanchez-Portal}, {Sturm}, {Tacconi}, {Valtchanov}, {Amblard},
  {Arumugam}, {Bethermin}, {Bock}, {Boselli}, {Buat}, {Burgarella},
  {Castro-Rodr{\'\i}guez}, {Cava}, {Chanial}, {Clements}, {Conley}, {Cooray},
  {Dowell}, {Dwek}, {Eales}, {Franceschini}, {Glenn}, {Griffin},
  {Hatziminaoglou}, {Ibar}, {Isaak}, {Ivison}, {Lagache}, {Levenson}, {Lu},
  {Madden}, {Maffei}, {Mainetti}, {Nguyen}, {O'Halloran}, {Page}, {Panuzzo},
  {Papageorgiou}, {Pearson}, {P{\'e}rez-Fournon}, {Pohlen}, {Rigopoulou},
  {Rowan-Robinson}, {Schulz}, {Scott}, {Seymour}, {Shupe}, {Smith}, {Stevens},
  {Symeonidis}, {Trichas}, {Tugwell}, {Vigroux}, {Wang}, {Wright}, {Xu},
  {Zemcov}, {Bardelli}, {Carollo}, {Contini}, {Le F{\'e}vre}, {Lilly},
  {Mainieri}, {Renzini}, {Scodeggio}, \& {Zucca}}]{Gruppioni2013}
{Gruppioni}, C., {Pozzi}, F., {Rodighiero}, G., {et~al.} 2013, \mnras, 432, 23

\bibitem[{{Harrington} {et~al.}(2019){Harrington}, {Vishwas}, {Wei{\ss}},
  {Magnelli}, {Grassitelli}, {Zaja{\v{c}}ek}, {Jim{\'e}nez-Andrade}, {Leung},
  {Bertoldi}, {Romano-D{\'\i}az}, {Frayer}, {Kamieneski}, {Riechers}, {Stacey},
  {Yun}, \& {Wang}}]{Harrington2019}
{Harrington}, K.~C., {Vishwas}, A., {Wei{\ss}}, A., {et~al.} 2019, \mnras, 488,
  1489

\bibitem[{{Harrington} {et~al.}(2021){Harrington}, {Weiss}, {Yun}, {Magnelli},
  {Sharon}, {Leung}, {Vishwas}, {Wang}, {Frayer}, {Jim{\'e}nez-Andrade}, {Liu},
  {Garc{\'\i}a}, {Romano-D{\'\i}az}, {Frye}, {Jarugula}, {B{\u{a}}descu},
  {Berman}, {Dannerbauer}, {D{\'\i}az-S{\'a}nchez}, {Grassitelli},
  {Kamieneski}, {Kim}, {Kirkpatrick}, {Lowenthal}, {Messias}, {Puschnig},
  {Stacey}, {Torne}, \& {Bertoldi}}]{Harrington2020}
{Harrington}, K.~C., {Weiss}, A., {Yun}, M.~S., {et~al.} 2021, \apj, 908, 95

\bibitem[{{Harrington} {et~al.}(2016){Harrington}, {Yun}, {Cybulski}, {Wilson},
  {Aretxaga}, {Chavez}, {De la Luz}, {Erickson}, {Ferrusca}, {Gallup},
  {Hughes}, {Monta{\~n}a}, {Narayanan}, {S{\'a}nchez-Arg{\"u}elles},
  {Schloerb}, {Souccar}, {Terlevich}, {Terlevich}, {Zeballos}, \&
  {Zavala}}]{Harrington2016}
{Harrington}, K.~C., {Yun}, M.~S., {Cybulski}, R., {et~al.} 2016, \mnras, 458,
  4383

\bibitem[{{Harrington} {et~al.}(2018){Harrington}, {Yun}, {Magnelli}, {Frayer},
  {Karim}, {Wei{\ss}}, {Riechers}, {Jim{\'e}nez-Andrade}, {Berman},
  {Lowenthal}, \& {Bertoldi}}]{Harrington2018}
{Harrington}, K.~C., {Yun}, M.~S., {Magnelli}, B., {et~al.} 2018, \mnras, 474,
  3866

\bibitem[{{Herranz} {et~al.}(2013){Herranz}, {Gonz{\'a}lez-Nuevo}, {Clements},
  {De Zotti}, {Lopez-Caniego}, {Lapi}, {Rodighiero}, {Danese}, {Fu}, {Cooray},
  {Baes}, {Bendo}, {Bonavera}, {Carrera}, {Dole}, {Eales}, {Ivison}, {Jarvis},
  {Lagache}, {Massardi}, {Micha{\l}owski}, {Negrello}, {Rigby}, {Scott},
  {Valiante}, {Valtchanov}, {Van der Werf}, {Auld}, {Buttiglione}, {Dariush},
  {Dunne}, {Hopwood}, {Hoyos}, {Ibar}, \& {Maddox}}]{Herranz2013}
{Herranz}, D., {Gonz{\'a}lez-Nuevo}, J., {Clements}, D.~L., {et~al.} 2013,
  \aap, 549, A31

\bibitem[{{Hezaveh} \& {Holder}(2011)}]{HezavehHolder2011}
{Hezaveh}, Y.~D. \& {Holder}, G.~P. 2011, \apj, 734, 52

\bibitem[{{Hilbert} {et~al.}(2008){Hilbert}, {White}, {Hartlap}, \&
  {Schneider}}]{Hilbert2008}
{Hilbert}, S., {White}, S. D.~M., {Hartlap}, J., \& {Schneider}, P. 2008,
  \mnras, 386, 1845

\bibitem[{{Hilton} {et~al.}(2021){Hilton}, {Sif{\'o}n}, {Naess},
  {Madhavacheril}, {Oguri}, {Rozo}, {Rykoff}, {Abbott}, {Adhikari}, {Aguena},
  {Aiola}, {Allam}, {Amodeo}, {Amon}, {Annis}, {Ansarinejad}, {Aros-Bunster},
  {Austermann}, {Avila}, {Bacon}, {Battaglia}, {Beall}, {Becker}, {Bernstein},
  {Bertin}, {Bhandarkar}, {Bhargava}, {Bond}, {Brooks}, {Burke}, {Calabrese},
  {Carrasco Kind}, {Carretero}, {Choi}, {Choi}, {Conselice}, {da Costa},
  {Costanzi}, {Crichton}, {Crowley}, {D{\"u}nner}, {Denison}, {Devlin},
  {Dicker}, {Diehl}, {Dietrich}, {Doel}, {Duff}, {Duivenvoorden}, {Dunkley},
  {Everett}, {Ferraro}, {Ferrero}, {Fert{\'e}}, {Flaugher}, {Frieman},
  {Gallardo}, {Garc{\'\i}a-Bellido}, {Gaztanaga}, {Gerdes}, {Giles}, {Golec},
  {Gralla}, {Grandis}, {Gruen}, {Gruendl}, {Gschwend}, {Gutierrez}, {Han},
  {Hartley}, {Hasselfield}, {Hill}, {Hilton}, {Hincks}, {Hinton}, {Ho},
  {Honscheid}, {Hoyle}, {Hubmayr}, {Huffenberger}, {Hughes}, {Jaelani}, {Jain},
  {James}, {Jeltema}, {Kent}, {Knowles}, {Koopman}, {Kuehn}, {Lahav}, {Lima},
  {Lin}, {Lokken}, {Loubser}, {MacCrann}, {Maia}, {Marriage}, {Martin},
  {McMahon}, {Melchior}, {Menanteau}, {Miquel}, {Miyatake}, {Moodley},
  {Morgan}, {Mroczkowski}, {Nati}, {Newburgh}, {Niemack}, {Nishizawa},
  {Ogando}, {Orlowski-Scherer}, {Page}, {Palmese}, {Partridge},
  {Paz-Chinch{\'o}n}, {Phakathi}, {Plazas}, {Robertson}, {Romer}, {Carnero
  Rosell}, {Salatino}, {Sanchez}, {Schaan}, {Schillaci}, {Sehgal}, {Serrano},
  {Shin}, {Simon}, {Smith}, {Soares-Santos}, {Spergel}, {Staggs}, {Storer},
  {Suchyta}, {Swanson}, {Tarle}, {Thomas}, {To}, {Trac}, {Ullom}, {Vale}, {Van
  Lanen}, {Vavagiakis}, {De Vicente}, {Wilkinson}, {Wollack}, {Xu}, \&
  {Zhang}}]{Hilton2021}
{Hilton}, M., {Sif{\'o}n}, C., {Naess}, S., {et~al.} 2021, \apjs, 253, 3

\bibitem[{{Hodge} {et~al.}(2016){Hodge}, {Swinbank}, {Simpson}, {Smail},
  {Walter}, {Alexander}, {Bertoldi}, {Biggs}, {Brandt}, {Chapman}, {Chen},
  {Coppin}, {Cox}, {Dannerbauer}, {Edge}, {Greve}, {Ivison}, {Karim},
  {Knudsen}, {Menten}, {Rix}, {Schinnerer}, {Wardlow}, {Weiss}, \& {van der
  Werf}}]{Hodge2016}
{Hodge}, J.~A., {Swinbank}, A.~M., {Simpson}, J.~M., {et~al.} 2016, \apj, 833,
  103

\bibitem[{{Ikarashi} {et~al.}(2017){Ikarashi}, {Caputi}, {Ohta}, {Ivison},
  {Lagos}, {Bisigello}, {Hatsukade}, {Aretxaga}, {Dunlop}, {Hughes}, {Iono},
  {Izumi}, {Kashikawa}, {Koyama}, {Kawabe}, {Kohno}, {Motohara}, {Nakanishi},
  {Tamura}, {Umehata}, {Wilson}, {Yabe}, \& {Yun}}]{Ikarashi2017}
{Ikarashi}, S., {Caputi}, K.~I., {Ohta}, K., {et~al.} 2017, \apjl, 849, L36

\bibitem[{{Jarrett} {et~al.}(2017){Jarrett}, {Cluver}, {Magoulas}, {Bilicki},
  {Alpaslan}, {Bland-Hawthorn}, {Brough}, {Brown}, {Croom}, {Driver},
  {Holwerda}, {Hopkins}, {Loveday}, {Norberg}, {Peacock}, {Popescu}, {Sadler},
  {Taylor}, {Tuffs}, \& {Wang}}]{Jarrett2017}
{Jarrett}, T.~H., {Cluver}, M.~E., {Magoulas}, C., {et~al.} 2017, \apj, 836,
  182

\bibitem[{{Jones} {et~al.}(2019){Jones}, {Maiolino}, {Caselli}, \&
  {Carniani}}]{Jones2019}
{Jones}, G.~C., {Maiolino}, R., {Caselli}, P., \& {Carniani}, S. 2019, \aap,
  632, L7

\bibitem[{{King} \& {Pounds}(2015)}]{KingPounds2015}
{King}, A. \& {Pounds}, K. 2015, \araa, 53, 115

\bibitem[{{Lagache} {et~al.}(2005){Lagache}, {Puget}, \& {Dole}}]{Lagache2005}
{Lagache}, G., {Puget}, J.-L., \& {Dole}, H. 2005, \araa, 43, 727

\bibitem[{{Lapi} {et~al.}(2012){Lapi}, {Negrello}, {Gonz{\'a}lez-Nuevo}, {Cai},
  {De Zotti}, \& {Danese}}]{Lapi2012}
{Lapi}, A., {Negrello}, M., {Gonz{\'a}lez-Nuevo}, J., {et~al.} 2012, \apj, 755,
  46

\bibitem[{{Lima} {et~al.}(2010){Lima}, {Jain}, {Devlin}, \&
  {Aguirre}}]{Lima2010}
{Lima}, M., {Jain}, B., {Devlin}, M., \& {Aguirre}, J. 2010, \apjl, 717, L31

\bibitem[{{Maddox} {et~al.}(2018){Maddox}, {Valiante}, {Cigan}, {Dunne},
  {Eales}, {Smith}, {Dye}, {Furlanetto}, {Ibar}, {de Zotti}, {Millard},
  {Bourne}, {Gomez}, {Ivison}, {Scott}, \& {Valtchanov}}]{Maddox2018}
{Maddox}, S.~J., {Valiante}, E., {Cigan}, P., {et~al.} 2018, \apjs, 236, 30

\bibitem[{{Merloni} {et~al.}(2012){Merloni}, {Predehl}, {Becker},
  {B{\"o}hringer}, {Boller}, {Brunner}, {Brusa}, {Dennerl}, {Freyberg},
  {Friedrich}, {Georgakakis}, {Haberl}, {Hasinger}, {Meidinger}, {Mohr},
  {Nandra}, {Rau}, {Reiprich}, {Robrade}, {Salvato}, {Santangelo}, {Sasaki},
  {Schwope}, {Wilms}, \& {German eROSITA Consortium}}]{Merloni2012}
{Merloni}, A., {Predehl}, P., {Becker}, W., {et~al.} 2012, arXiv e-prints,
  arXiv:1209.3114

\bibitem[{{Miville-Desch{\^e}nes} \& {Lagache}(2005)}]{MivilleDeschenes2005}
{Miville-Desch{\^e}nes}, M.-A. \& {Lagache}, G. 2005, \apjs, 157, 302

\bibitem[{{Mocanu} {et~al.}(2013){Mocanu}, {Crawford}, {Vieira}, {Aird},
  {Aravena}, {Austermann}, {Benson}, {B{\'e}thermin}, {Bleem}, {Bothwell},
  {Carlstrom}, {Chang}, {Chapman}, {Cho}, {Crites}, {de Haan}, {Dobbs},
  {Everett}, {George}, {Halverson}, {Harrington}, {Hezaveh}, {Holder},
  {Holzapfel}, {Hoover}, {Hrubes}, {Keisler}, {Knox}, {Lee}, {Leitch},
  {Lueker}, {Luong-Van}, {Marrone}, {McMahon}, {Mehl}, {Meyer}, {Mohr},
  {Montroy}, {Natoli}, {Padin}, {Plagge}, {Pryke}, {Rest}, {Reichardt}, {Ruhl},
  {Sayre}, {Schaffer}, {Shirokoff}, {Spieler}, {Spilker}, {Stalder},
  {Staniszewski}, {Stark}, {Story}, {Switzer}, {Vanderlinde}, \&
  {Williamson}}]{Mocanu2013}
{Mocanu}, L.~M., {Crawford}, T.~M., {Vieira}, J.~D., {et~al.} 2013, \apj, 779,
  61

\bibitem[{{Nayyeri} {et~al.}(2016){Nayyeri}, {Keele}, {Cooray}, {Riechers},
  {Ivison}, {Harris}, {Frayer}, {Baker}, {Chapman}, {Eales}, {Farrah}, {Fu},
  {Marchetti}, {Marques-Chaves}, {Martinez-Navajas}, {Oliver}, {Omont},
  {Perez-Fournon}, {Scott}, {Vaccari}, {Vieira}, {Viero}, {Wang}, \&
  {Wardlow}}]{Nayyeri2016}
{Nayyeri}, H., {Keele}, M., {Cooray}, A., {et~al.} 2016, \apj, 823, 17

\bibitem[{{Negrello} {et~al.}(2017){Negrello}, {Amber}, {Amvrosiadis}, {Cai},
  {Lapi}, {Gonzalez-Nuevo}, {De Zotti}, {Furlanetto}, {Maddox}, {Allen},
  {Bakx}, {Bussmann}, {Cooray}, {Covone}, {Danese}, {Dannerbauer}, {Fu},
  {Greenslade}, {Gurwell}, {Hopwood}, {Koopmans}, {Napolitano}, {Nayyeri},
  {Omont}, {Petrillo}, {Riechers}, {Serjeant}, {Tortora}, {Valiante}, {Verdoes
  Kleijn}, {Vernardos}, {Wardlow}, {Baes}, {Baker}, {Bourne}, {Clements},
  {Crawford}, {Dye}, {Dunne}, {Eales}, {Ivison}, {Marchetti}, {Micha{\l}owski},
  {Smith}, {Vaccari}, \& {van der Werf}}]{Negrello2017lensed}
{Negrello}, M., {Amber}, S., {Amvrosiadis}, A., {et~al.} 2017, \mnras, 465,
  3558

\bibitem[{{Negrello} {et~al.}(2013){Negrello}, {Clemens}, {Gonzalez-Nuevo}, {De
  Zotti}, {Bonavera}, {Cosco}, {Guarese}, {Boaretto}, {Serjeant}, {Toffolatti},
  {Lapi}, {Bethermin}, {Castex}, {Clements}, {Delabrouille}, {Dole},
  {Franceschini}, {Mandolesi}, {Marchetti}, {Partridge}, \&
  {Sajina}}]{Negrello2013}
{Negrello}, M., {Clemens}, M., {Gonzalez-Nuevo}, J., {et~al.} 2013, \mnras,
  429, 1309

\bibitem[{{Negrello} {et~al.}(2010){Negrello}, {Hopwood}, {De Zotti}, {Cooray},
  {Verma}, {Bock}, {Frayer}, {Gurwell}, {Omont}, {Neri}, {Dannerbauer},
  {Leeuw}, {Barton}, {Cooke}, {Kim}, {da Cunha}, {Rodighiero}, {Cox},
  {Bonfield}, {Jarvis}, {Serjeant}, {Ivison}, {Dye}, {Aretxaga}, {Hughes},
  {Ibar}, {Bertoldi}, {Valtchanov}, {Eales}, {Dunne}, {Driver}, {Auld},
  {Buttiglione}, {Cava}, {Grady}, {Clements}, {Dariush}, {Fritz}, {Hill},
  {Hornbeck}, {Kelvin}, {Lagache}, {Lopez-Caniego}, {Gonzalez-Nuevo}, {Maddox},
  {Pascale}, {Pohlen}, {Rigby}, {Robotham}, {Simpson}, {Smith}, {Temi},
  {Thompson}, {Woodgate}, {York}, {Aguirre}, {Beelen}, {Blain}, {Baker},
  {Birkinshaw}, {Blundell}, {Bradford}, {Burgarella}, {Danese}, {Dunlop},
  {Fleuren}, {Glenn}, {Harris}, {Kamenetzky}, {Lupu}, {Maddalena}, {Madore},
  {Maloney}, {Matsuhara}, {Micha{\l}owski}, {Murphy}, {Naylor}, {Nguyen},
  {Popescu}, {Rawlings}, {Rigopoulou}, {Scott}, {Scott}, {Seibert}, {Smail},
  {Tuffs}, {Vieira}, {van der Werf}, \& {Zmuidzinas}}]{Negrello2010}
{Negrello}, M., {Hopwood}, R., {De Zotti}, G., {et~al.} 2010, Science, 330, 800

\bibitem[{{Negrello} {et~al.}(2007){Negrello}, {Perrotta},
  {Gonz{\'a}lez-Nuevo}, {Silva}, {de Zotti}, {Granato}, {Baccigalupi}, \&
  {Danese}}]{Negrello2007}
{Negrello}, M., {Perrotta}, F., {Gonz{\'a}lez-Nuevo}, J., {et~al.} 2007,
  \mnras, 377, 1557

\bibitem[{{Nesvadba} {et~al.}(2016){Nesvadba}, {Kneissl}, {Ca{\~n}ameras},
  {Boone}, {Falgarone}, {Frye}, {Gerin}, {Koenig}, {Lagache}, {Le Floc'h},
  {Malhotra}, \& {Scott}}]{Nesvadba2016}
{Nesvadba}, N., {Kneissl}, R., {Ca{\~n}ameras}, R., {et~al.} 2016, \aap, 593,
  L2

\bibitem[{{Nesvadba} {et~al.}(2019){Nesvadba}, {Ca{\~n}ameras}, {Kneissl},
  {Koenig}, {Yang}, {Le Floc'h}, {Omont}, \& {Scott}}]{Nesvadba2019}
{Nesvadba}, N.~P.~H., {Ca{\~n}ameras}, R., {Kneissl}, R., {et~al.} 2019, \aap,
  624, A23

\bibitem[{{Oliver} {et~al.}(2012){Oliver}, {Bock}, {Altieri}, {Amblard},
  {Arumugam}, {Aussel}, {Babbedge}, {Beelen}, {B{\'e}thermin}, {Blain},
  {Boselli}, {Bridge}, {Brisbin}, {Buat}, {Burgarella},
  {Castro-Rodr{\'\i}guez}, {Cava}, {Chanial}, {Cirasuolo}, {Clements},
  {Conley}, {Conversi}, {Cooray}, {Dowell}, {Dubois}, {Dwek}, {Dye}, {Eales},
  {Elbaz}, {Farrah}, {Feltre}, {Ferrero}, {Fiolet}, {Fox}, {Franceschini},
  {Gear}, {Giovannoli}, {Glenn}, {Gong}, {Gonz{\'a}lez Solares}, {Griffin},
  {Halpern}, {Harwit}, {Hatziminaoglou}, {Heinis}, {Hurley}, {Hwang}, {Hyde},
  {Ibar}, {Ilbert}, {Isaak}, {Ivison}, {Lagache}, {Le Floc'h}, {Levenson},
  {Faro}, {Lu}, {Madden}, {Maffei}, {Magdis}, {Mainetti}, {Marchetti},
  {Marsden}, {Marshall}, {Mortier}, {Nguyen}, {O'Halloran}, {Omont}, {Page},
  {Panuzzo}, {Papageorgiou}, {Patel}, {Pearson}, {P{\'e}rez-Fournon}, {Pohlen},
  {Rawlings}, {Raymond}, {Rigopoulou}, {Riguccini}, {Rizzo}, {Rodighiero},
  {Roseboom}, {Rowan-Robinson}, {S{\'a}nchez Portal}, {Schulz}, {Scott},
  {Seymour}, {Shupe}, {Smith}, {Stevens}, {Symeonidis}, {Trichas}, {Tugwell},
  {Vaccari}, {Valtchanov}, {Vieira}, {Viero}, {Vigroux}, {Wang}, {Ward},
  {Wardlow}, {Wright}, {Xu}, \& {Zemcov}}]{Oliver2012}
{Oliver}, S.~J., {Bock}, J., {Altieri}, B., {et~al.} 2012, \mnras, 424, 1614

\bibitem[{{Perrotta} {et~al.}(2002){Perrotta}, {Baccigalupi}, {Bartelmann}, {De
  Zotti}, \& {Granato}}]{Perrotta2002}
{Perrotta}, F., {Baccigalupi}, C., {Bartelmann}, M., {De Zotti}, G., \&
  {Granato}, G.~L. 2002, \mnras, 329, 445

\bibitem[{{Perrotta} {et~al.}(2003){Perrotta}, {Magliocchetti}, {Baccigalupi},
  {Bartelmann}, {De Zotti}, {Granato}, {Silva}, \& {Danese}}]{Perrotta2003}
{Perrotta}, F., {Magliocchetti}, M., {Baccigalupi}, C., {et~al.} 2003, \mnras,
  338, 623

\bibitem[{{Piffaretti} {et~al.}(2011){Piffaretti}, {Arnaud}, {Pratt},
  {Pointecouteau}, \& {Melin}}]{Piffaretti2011}
{Piffaretti}, R., {Arnaud}, M., {Pratt}, G.~W., {Pointecouteau}, E., \&
  {Melin}, J.~B. 2011, \aap, 534, A109

\bibitem[{{Planck Collaboration} {et~al.}(2020){Planck Collaboration},
  {Aghanim}, {Akrami}, {Arroja}, {Ashdown}, {Aumont}, {Baccigalupi},
  {Ballardini}, {Banday}, {Barreiro}, {Bartolo}, {Basak}, {Battye}, {Benabed},
  {Bernard}, {Bersanelli}, {Bielewicz}, {Bock}, {Bond}, {Borrill}, {Bouchet},
  {Boulanger}, {Bucher}, {Burigana}, {Butler}, {Calabrese}, {Cardoso},
  {Carron}, {Casaponsa}, {Challinor}, {Chiang}, {Colombo}, {Combet},
  {Contreras}, {Crill}, {Cuttaia}, {de Bernardis}, {de Zotti}, {Delabrouille},
  {Delouis}, {D{\'e}sert}, {Di Valentino}, {Dickinson}, {Diego}, {Donzelli},
  {Dor{\'e}}, {Douspis}, {Ducout}, {Dupac}, {Efstathiou}, {Elsner},
  {En{\ss}lin}, {Eriksen}, {Falgarone}, {Fantaye}, {Fergusson},
  {Fernandez-Cobos}, {Finelli}, {Forastieri}, {Frailis}, {Franceschi},
  {Frolov}, {Galeotta}, {Galli}, {Ganga}, {G{\'e}nova-Santos}, {Gerbino},
  {Ghosh}, {Gonz{\'a}lez-Nuevo}, {G{\'o}rski}, {Gratton}, {Gruppuso},
  {Gudmundsson}, {Hamann}, {Handley}, {Hansen}, {Helou}, {Herranz},
  {Hildebrandt}, {Hivon}, {Huang}, {Jaffe}, {Jones}, {Karakci}, {Keih{\"a}nen},
  {Keskitalo}, {Kiiveri}, {Kim}, {Kisner}, {Knox}, {Krachmalnicoff}, {Kunz},
  {Kurki-Suonio}, {Lagache}, {Lamarre}, {Langer}, {Lasenby}, {Lattanzi},
  {Lawrence}, {Le Jeune}, {Leahy}, {Lesgourgues}, {Levrier}, {Lewis},
  {Liguori}, {Lilje}, {Lilley}, {Lindholm}, {L{\'o}pez-Caniego}, {Lubin}, {Ma},
  {Mac{\'\i}as-P{\'e}rez}, {Maggio}, {Maino}, {Mandolesi}, {Mangilli},
  {Marcos-Caballero}, {Maris}, {Martin}, {Martinelli},
  {Mart{\'\i}nez-Gonz{\'a}lez}, {Matarrese}, {Mauri}, {McEwen}, {Meerburg},
  {Meinhold}, {Melchiorri}, {Mennella}, {Migliaccio}, {Millea}, {Mitra},
  {Miville-Desch{\^e}nes}, {Molinari}, {Moneti}, {Montier}, {Morgante}, {Moss},
  {Mottet}, {M{\"u}nchmeyer}, {Natoli}, {N{\o}rgaard-Nielsen}, {Oxborrow},
  {Pagano}, {Paoletti}, {Partridge}, {Patanchon}, {Pearson}, {Peel}, {Peiris},
  {Perrotta}, {Pettorino}, {Piacentini}, {Polastri}, {Polenta}, {Puget},
  {Rachen}, {Reinecke}, {Remazeilles}, {Renault}, {Renzi}, {Rocha}, {Rosset},
  {Roudier}, {Rubi{\~n}o-Mart{\'\i}n}, {Ruiz-Granados}, {Salvati}, {Sandri},
  {Savelainen}, {Scott}, {Shellard}, {Shiraishi}, {Sirignano}, {Sirri},
  {Spencer}, {Sunyaev}, {Suur-Uski}, {Tauber}, {Tavagnacco}, {Tenti},
  {Terenzi}, {Toffolatti}, {Tomasi}, {Trombetti}, {Valiviita}, {Van Tent},
  {Vibert}, {Vielva}, {Villa}, {Vittorio}, {Wandelt}, {Wehus}, {White},
  {White}, {Zacchei}, \& {Zonca}}]{PlanckCollaboration2020overview}
{Planck Collaboration}, {Aghanim}, N., {Akrami}, Y., {et~al.} 2020, \aap, 641,
  A1

\bibitem[{{Planck Collaboration
  I}(2016)}]{PlanckCollaboration2016data_products}
{Planck Collaboration I}. 2016, \aap, 594, A1

\bibitem[{{Planck Collaboration Int. LIV}(2018)}]{PlanckCollaboration2018PCNT}
{Planck Collaboration Int. LIV}. 2018, \aap, 619, A94

\bibitem[{{Planck Collaboration Int. LV}(2020)}]{PlanckCollaboration2020BEEP}
{Planck Collaboration Int. LV}. 2020, \aap, 644, A99

\bibitem[{{Planck Collaboration Int.
  VII}(2013)}]{PlanckCollaboration2013counts}
{Planck Collaboration Int. VII}. 2013, \aap, 550, A133

\bibitem[{{Planck Collaboration Int.
  XXVII}(2015)}]{PlanckCollaboration2015clumpsHerschel}
{Planck Collaboration Int. XXVII}. 2015, \aap, 582, A30

\bibitem[{{Planck Collaboration Int.
  XXXIX}(2016)}]{PlanckCollaboration2016highz}
{Planck Collaboration Int. XXXIX}. 2016, \aap, 596, A100

\bibitem[{{Planck Collaboration VII}(2011)}]{ERCSC2011}
{Planck Collaboration VII}. 2011, \aap, 536, A7

\bibitem[{{Planck Collaboration XV}(2014)}]{PlanckCollaboration2014}
{Planck Collaboration XV}. 2014, \aap, 571, A15

\bibitem[{{Planck Collaboration XXVI}(2016)}]{PCCS2}
{Planck Collaboration XXVI}. 2016, \aap, 594, A26

\bibitem[{{Planck Collaboration XXVII}(2016)}]{PlanckCollaboration2016SZ}
{Planck Collaboration XXVII}. 2016, \aap, 594, A27

\bibitem[{{Planck Collaboration
  XXVIII}(2016)}]{PlanckCollaboration2016cold_cores}
{Planck Collaboration XXVIII}. 2016, \aap, 594, A28

\bibitem[{{Reuter} {et~al.}(2020){Reuter}, {Vieira}, {Spilker}, {Weiss},
  {Aravena}, {Archipley}, {B{\'e}thermin}, {Chapman}, {De Breuck}, {Dong},
  {Everett}, {Fu}, {Greve}, {Hayward}, {Hill}, {Hezaveh}, {Jarugula}, {Litke},
  {Malkan}, {Marrone}, {Narayanan}, {Phadke}, {Stark}, \&
  {Strandet}}]{Reuter2020}
{Reuter}, C., {Vieira}, J.~D., {Spilker}, J.~S., {et~al.} 2020, \apj, 902, 78

\bibitem[{{Robertson} {et~al.}(2020){Robertson}, {Smith}, {Massey}, {Eke},
  {Jauzac}, {Bianconi}, \& {Ryczanowski}}]{Robertson2020}
{Robertson}, A., {Smith}, G.~P., {Massey}, R., {et~al.} 2020, \mnras, 495, 3727

\bibitem[{{Rowan-Robinson} {et~al.}(2018){Rowan-Robinson}, {Wang}, {Farrah},
  {Rigopoulou}, {Gruppioni}, {Vaccari}, {Marchetti}, {Clements}, \&
  {Pearson}}]{RowanRobinson2018}
{Rowan-Robinson}, M., {Wang}, L., {Farrah}, D., {et~al.} 2018, \aap, 619, A169

\bibitem[{{Shibuya} {et~al.}(2015){Shibuya}, {Ouchi}, \&
  {Harikane}}]{Shibuya2015}
{Shibuya}, T., {Ouchi}, M., \& {Harikane}, Y. 2015, \apjs, 219, 15

\bibitem[{{Shibuya} {et~al.}(2019){Shibuya}, {Ouchi}, {Harikane}, \&
  {Nakajima}}]{Shibuya2019}
{Shibuya}, T., {Ouchi}, M., {Harikane}, Y., \& {Nakajima}, K. 2019, \apj, 871,
  164

\bibitem[{{Shu} {et~al.}(2016){Shu}, {Bolton}, {Kochanek}, {Oguri},
  {P{\'e}rez-Fournon}, {Zheng}, {Mao}, {Montero-Dorta}, {Brownstein},
  {Marques-Chaves}, \& {M{\'e}nard}}]{Shu2016}
{Shu}, Y., {Bolton}, A.~S., {Kochanek}, C.~S., {et~al.} 2016, \apj, 824, 86

\bibitem[{{Silk} \& {Mamon}(2012)}]{SilkMamon2012}
{Silk}, J. \& {Mamon}, G.~A. 2012, Research in Astronomy and Astrophysics, 12,
  917

\bibitem[{{Somerville} \& {Dav{\'e}}(2015)}]{SomervilleDave2015}
{Somerville}, R.~S. \& {Dav{\'e}}, R. 2015, \araa, 53, 51

\bibitem[{{Spilker} {et~al.}(2018){Spilker}, {Aravena}, {B{\'e}thermin},
  {Chapman}, {Chen}, {Cunningham}, {De Breuck}, {Dong}, {Gonzalez}, {Hayward},
  {Hezaveh}, {Litke}, {Ma}, {Malkan}, {Marrone}, {Miller}, {Morningstar},
  {Narayanan}, {Phadke}, {Sreevani}, {Stark}, {Vieira}, \&
  {Wei{\ss}}}]{Spilker2018}
{Spilker}, J.~S., {Aravena}, M., {B{\'e}thermin}, M., {et~al.} 2018, Science,
  361, 1016

\bibitem[{{Spilker} {et~al.}(2016){Spilker}, {Marrone}, {Aravena},
  {B{\'e}thermin}, {Bothwell}, {Carlstrom}, {Chapman}, {Crawford}, {de Breuck},
  {Fassnacht}, {Gonzalez}, {Greve}, {Hezaveh}, {Litke}, {Ma}, {Malkan},
  {Rotermund}, {Strandet}, {Vieira}, {Weiss}, \& {Welikala}}]{Spilker2016}
{Spilker}, J.~S., {Marrone}, D.~P., {Aravena}, M., {et~al.} 2016, \apj, 826,
  112

\bibitem[{{Spilker} {et~al.}(2020){Spilker}, {Phadke}, {Aravena},
  {B{\'e}thermin}, {Chapman}, {Dong}, {Gonzalez}, {Hayward}, {Hezaveh},
  {Jarugula}, {Litke}, {Malkan}, {Marrone}, {Narayanan}, {Reuter}, {Vieira}, \&
  {Weiss}}]{Spilker2020}
{Spilker}, J.~S., {Phadke}, K.~A., {Aravena}, M., {et~al.} 2020, \apj, 905, 85

\bibitem[{{Sun} {et~al.}(2021){Sun}, {Egami}, {Rawle}, {Walth}, {Smail},
  {Dessauges-Zavadsky}, {P{\'e}rez-Gonz{\'a}lez}, {Richard}, {Combes},
  {Ebeling}, {Pell{\'o}}, {Van der Werf}, {Altieri}, {Boone}, {Cava},
  {Chapman}, {Cl{\'e}ment}, {Finoguenov}, {Nakajima}, {Rujopakarn}, {Schaerer},
  \& {Valtchanov}}]{Sun2021}
{Sun}, F., {Egami}, E., {Rawle}, T.~D., {et~al.} 2021, \apj, 908, 192

\bibitem[{{Swinbank} {et~al.}(2010){Swinbank}, {Smail}, {Longmore}, {Harris},
  {Baker}, {De Breuck}, {Richard}, {Edge}, {Ivison}, {Blundell}, {Coppin},
  {Cox}, {Gurwell}, {Hainline}, {Krips}, {Lundgren}, {Neri}, {Siana},
  {Siringo}, {Stark}, {Wilner}, \& {Younger}}]{Swinbank2010}
{Swinbank}, A.~M., {Smail}, I., {Longmore}, S., {et~al.} 2010, \nat, 464, 733

\bibitem[{{Talbot} {et~al.}(2021){Talbot}, {Brownstein}, {Dawson}, {Kneib}, \&
  {Bautista}}]{Talbot2020}
{Talbot}, M.~S., {Brownstein}, J.~R., {Dawson}, K.~S., {Kneib}, J.-P., \&
  {Bautista}, J. 2021, \mnras, 502, 4617

\bibitem[{{Taylor}(2005)}]{TOPCAT2005}
{Taylor}, M.~B. 2005, in Astronomical Society of the Pacific Conference Series,
  Vol. 347, Astronomical Data Analysis Software and Systems XIV, ed.
  P.~{Shopbell}, M.~{Britton}, \& R.~{Ebert}, 29

\bibitem[{{Tucci} {et~al.}(2011){Tucci}, {Toffolatti}, {de Zotti}, \&
  {Mart{\'\i}nez-Gonz{\'a}lez}}]{Tucci2011}
{Tucci}, M., {Toffolatti}, L., {de Zotti}, G., \& {Mart{\'\i}nez-Gonz{\'a}lez},
  E. 2011, \aap, 533, A57

\bibitem[{{Veilleux} {et~al.}(2020){Veilleux}, {Maiolino}, {Bolatto}, \&
  {Aalto}}]{Veilleux2020}
{Veilleux}, S., {Maiolino}, R., {Bolatto}, A.~D., \& {Aalto}, S. 2020, \aapr,
  28, 2

\bibitem[{{Vieira} {et~al.}(2010){Vieira}, {Crawford}, {Switzer}, {Ade},
  {Aird}, {Ashby}, {Benson}, {Bleem}, {Brodwin}, {Carlstrom}, {Chang}, {Cho},
  {Crites}, {de Haan}, {Dobbs}, {Everett}, {George}, {Gladders}, {Hall},
  {Halverson}, {High}, {Holder}, {Holzapfel}, {Hrubes}, {Joy}, {Keisler},
  {Knox}, {Lee}, {Leitch}, {Lueker}, {Marrone}, {McIntyre}, {McMahon}, {Mehl},
  {Meyer}, {Mohr}, {Montroy}, {Padin}, {Plagge}, {Pryke}, {Reichardt}, {Ruhl},
  {Schaffer}, {Shaw}, {Shirokoff}, {Spieler}, {Stalder}, {Staniszewski},
  {Stark}, {Vanderlinde}, {Walsh}, {Williamson}, {Yang}, {Zahn}, \&
  {Zenteno}}]{Vieira2010}
{Vieira}, J.~D., {Crawford}, T.~M., {Switzer}, E.~R., {et~al.} 2010, \apj, 719,
  763

\bibitem[{{Viero} {et~al.}(2014){Viero}, {Asboth}, {Roseboom}, {Moncelsi},
  {Marsden}, {Mentuch Cooper}, {Zemcov}, {Addison}, {Baker}, {Beelen}, {Bock},
  {Bridge}, {Conley}, {Devlin}, {Dor{\'e}}, {Farrah}, {Finkelstein},
  {Font-Ribera}, {Geach}, {Gebhardt}, {Gill}, {Glenn}, {Hajian}, {Halpern},
  {Jogee}, {Kurczynski}, {Lapi}, {Negrello}, {Oliver}, {Papovich}, {Quadri},
  {Ross}, {Scott}, {Schulz}, {Somerville}, {Spergel}, {Vieira}, {Wang}, \&
  {Wechsler}}]{Viero2014}
{Viero}, M.~P., {Asboth}, V., {Roseboom}, I.~G., {et~al.} 2014, \apjs, 210, 22

\bibitem[{{Wardlow} {et~al.}(2013){Wardlow}, {Cooray}, {De Bernardis},
  {Amblard}, {Arumugam}, {Aussel}, {Baker}, {B{\'e}thermin}, {Blundell},
  {Bock}, {Boselli}, {Bridge}, {Buat}, {Burgarella}, {Bussmann},
  {Cabrera-Lavers}, {Calanog}, {Carpenter}, {Casey}, {Castro-Rodr{\'\i}guez},
  {Cava}, {Chanial}, {Chapin}, {Chapman}, {Clements}, {Conley}, {Cox},
  {Dowell}, {Dye}, {Eales}, {Farrah}, {Ferrero}, {Franceschini}, {Frayer},
  {Frazer}, {Fu}, {Gavazzi}, {Glenn}, {Gonz{\'a}lez Solares}, {Griffin},
  {Gurwell}, {Harris}, {Hatziminaoglou}, {Hopwood}, {Hyde}, {Ibar}, {Ivison},
  {Kim}, {Lagache}, {Levenson}, {Marchetti}, {Marsden}, {Martinez-Navajas},
  {Negrello}, {Neri}, {Nguyen}, {O'Halloran}, {Oliver}, {Omont}, {Page},
  {Panuzzo}, {Papageorgiou}, {Pearson}, {P{\'e}rez-Fournon}, {Pohlen},
  {Riechers}, {Rigopoulou}, {Roseboom}, {Rowan-Robinson}, {Schulz}, {Scott},
  {Scoville}, {Seymour}, {Shupe}, {Smith}, {Streblyanska}, {Strom},
  {Symeonidis}, {Trichas}, {Vaccari}, {Vieira}, {Viero}, {Wang}, {Xu}, {Yan},
  \& {Zemcov}}]{Wardlow2013}
{Wardlow}, J.~L., {Cooray}, A., {De Bernardis}, F., {et~al.} 2013, \apj, 762,
  59

\bibitem[{{Wei{\ss}} {et~al.}(2013){Wei{\ss}}, {De Breuck}, {Marrone},
  {Vieira}, {Aguirre}, {Aird}, {Aravena}, {Ashby}, {Bayliss}, {Benson},
  {B{\'e}thermin}, {Biggs}, {Bleem}, {Bock}, {Bothwell}, {Bradford}, {Brodwin},
  {Carlstrom}, {Chang}, {Chapman}, {Crawford}, {Crites}, {de Haan}, {Dobbs},
  {Downes}, {Fassnacht}, {George}, {Gladders}, {Gonzalez}, {Greve},
  {Halverson}, {Hezaveh}, {High}, {Holder}, {Holzapfel}, {Hoover}, {Hrubes},
  {Husband}, {Keisler}, {Lee}, {Leitch}, {Lueker}, {Luong-Van}, {Malkan},
  {McIntyre}, {McMahon}, {Mehl}, {Menten}, {Meyer}, {Murphy}, {Padin},
  {Plagge}, {Reichardt}, {Rest}, {Rosenman}, {Ruel}, {Ruhl}, {Schaffer},
  {Shirokoff}, {Spilker}, {Stalder}, {Staniszewski}, {Stark}, {Story},
  {Vanderlinde}, {Welikala}, \& {Williamson}}]{Weiss2013}
{Wei{\ss}}, A., {De Breuck}, C., {Marrone}, D.~P., {et~al.} 2013, \apj, 767, 88

\bibitem[{{Zavala} {et~al.}(2017){Zavala}, {Aretxaga}, {Geach}, {Hughes},
  {Birkinshaw}, {Chapin}, {Chapman}, {Chen}, {Clements}, {Dunlop}, {Farrah},
  {Ivison}, {Jenness}, {Micha{\l}owski}, {Robson}, {Scott}, {Simpson},
  {Spaans}, \& {van der Werf}}]{Zavala2017}
{Zavala}, J.~A., {Aretxaga}, I., {Geach}, J.~E., {et~al.} 2017, \mnras, 464,
  3369

\end{thebibliography}

\newpage
\appendix
\onecolumn
\section{Candidate strongly lensed galaxies selected at 545, 857 and 353\,GHz}
\begin{center}
\small 
\begin{longtable}{lccccccccc}
\caption{MTXF photometry of candidate strongly lensed galaxies selected at 545\,GHz. P stands for PCCS2; all sources, except for H\,G200.61+46.09 which comes from the PHz catalogue, were drawn from this catalogue. Flux densities and errors are in mJy, rounded to 1\,mJy. Equatorial coordinates, RA and DEC, J2000, in degrees. Notes: L for confirmed strongly lensed (Table\,\ref{tab:confirmed}); G for sources included in the \textit{Planck} Galactic cold clump catalogue \citep{PlanckCollaboration2016cold_cores}; H for sources included in the \textit{Planck} list of high-redshift source candidates \citep{PlanckCollaboration2016highz}; C for sources associated with galaxy clusters (see text); S for sources with counterparts within 3 arcmin in the SPT catalogue \citep{Everett2020}. }
\label{tab:MTXFcandidates545GHz}\\
\hline
  \multicolumn{1}{c}{\textit{Planck} name} &
  \multicolumn{1}{c}{RA} &
  \multicolumn{1}{c}{DEC} &
  \multicolumn{1}{c}{$S_{217}$}&
  \multicolumn{1}{c}{$S_{353}$}&
  \multicolumn{1}{c}{$S_{545}$}&
  \multicolumn{1}{c}{$S_{857}$}&
  \multicolumn{1}{c}{$S_{857}/S_{545}$}&
  \multicolumn{1}{c}{$S_{545}/S_{353}$}&
  \multicolumn{1}{c}{Note} \\
\hline
\endfirsthead

\multicolumn{10}{c}%
{{\bfseries \tablename\ \thetable{} -- continued from previous page}} \\
\hline
  \multicolumn{1}{c}{\textit{Planck} name} &
  \multicolumn{1}{c}{RA} &
  \multicolumn{1}{c}{DEC} &
  \multicolumn{1}{c}{$S_{217}$}&
  \multicolumn{1}{c}{$S_{353}$}&
  \multicolumn{1}{c}{$S_{545}$}&
  \multicolumn{1}{c}{$S_{857}$}&
  \multicolumn{1}{m{1cm}}{$S_{857}/S_{545}$}&
  \multicolumn{1}{m{1cm}}{$S_{545}/S_{353}$}&
  \multicolumn{1}{c}{Note} \\
\hline
\endhead

\hline \multicolumn{10}{r}{{Continued on next page}} \\ \hline
\endfoot

\hline \hline
\endlastfoot
P\,G107.82-45.46 & 1.8930 &  16.1184 & $ 135\pm 33$ & $ 567\pm  83$ & $ 922\pm 131$ & $1213\pm 334$ & 1.32 & 1.62 & G,C \\
P\,G318.60-68.18 & 4.2517 & -47.8788 & $  40\pm 26$ & $ 184\pm  65$ & $ 414\pm  94$ & $ 602\pm 151$ & 1.45 & 2.26 &  \\
P\,G116.05-45.04 & 7.7692 &  17.5577 & $ 119\pm 50$ & $ 393\pm 144$ & $1678\pm 417$ & $4470\pm1327$ & 2.66 & 4.27 &  \\
P\,G120.01-43.19 &10.6032 &  19.6206 & $  60\pm 36$ & $ 307\pm  81$ & $ 610\pm 136$ & $ 638\pm 339$ & 1.05 & 1.98 &  \\
P\,G124.53-28.73 & 14.5626 &  34.1135 & $  76\pm 37$ & $ 128\pm  96$ & $ 623\pm 156$ & $1514\pm 432$ & 2.43 & 4.85 &  \\
P\,G128.89-71.39 & 14.7807 &  -8.6103 & $  67\pm 35$ & $ 338\pm  79$ & $ 812\pm 142$ & $2209\pm 385$ & 2.72 & 2.40 &  \\
P\,G130.16-64.05 & 16.0164 &  -1.3606 & $  52\pm 32$ & $ 179\pm  80$ & $ 569\pm 122$ & $1269\pm 307$ & 2.23 & 3.17 &  \\
P\,G215.65-87.14 & 16.0636 & -27.2274 & $  49\pm 29$ & $ 113\pm  69$ & $ 553\pm 113$ & $ 615\pm 254$ & 1.11 & 4.91 & C \\
\hline
 \end{longtable}
\end{center}

\newpage
\begin{center}
\small \tabcolsep=0.07 cm
\begin{longtable}{lccccccccccc}

\caption{BeeP photometry of candidate strongly lensed galaxies selected at 857\,GHz. P and PE stand for PCCS2 and PCCS2E, respectively. Units and notes as in Table~\ref{tab:MTXFcandidates545GHz}.  The errors were derived from the 97.725\% confidence ($\simeq 2\,\sigma$) upper and lower boundaries of the distributions (S*H2SBU and S*L2SBU), taking the semi-difference between the boundaries and the central values to derive the $\simeq 1\,\sigma$ errors. SRS stands for SRCSIG, which is a measure of the source detection significance. SN stands for SNRR857, defined as the source average brightness divided the background standard deviation brightness;
$R_1\equiv S_{857}/S_{545}$, $R_2\equiv S_{857}/S_{3000}$.}
\label{tab:candidates857GHz}\\
\hline
  \multicolumn{1}{c}{\textit{Planck} name} &
  \multicolumn{1}{c}{RA} &
  \multicolumn{1}{c}{DEC} &
  \multicolumn{1}{c}{SRS} &
  \multicolumn{1}{c}{$S_{353}$}&
  \multicolumn{1}{c}{$S_{545}$}&
  \multicolumn{1}{c}{$S_{857}$}&
  \multicolumn{1}{c}{$S_{3000}$}&
  \multicolumn{1}{c}{SN} &
  \multicolumn{1}{c}{$R_1$}&
  \multicolumn{1}{c}{$R_2$}&
  \multicolumn{1}{c}{Note} \\
 \hline
\endfirsthead

\multicolumn{12}{c}%
{{\bfseries \tablename\ \thetable{} -- continued from previous page}} \\
\hline
  \multicolumn{1}{c}{\textit{Planck} name} &
  \multicolumn{1}{c}{RA} &
  \multicolumn{1}{c}{DEC} &
  \multicolumn{1}{c}{SRS} &
  \multicolumn{1}{c}{$S_{353}$}&
  \multicolumn{1}{c}{$S_{545}$}&
  \multicolumn{1}{c}{$S_{857}$}&
  \multicolumn{1}{c}{$S_{3000}$}&
  \multicolumn{1}{c}{SN} &
  \multicolumn{1}{c}{$R_1$}&
  \multicolumn{1}{c}{$R_2$}&
  \multicolumn{1}{c}{Note} \\
\hline
\endhead

\hline \multicolumn{12}{r}{{Continued on next page}} \\ \hline
\endfoot

\hline \hline
\endlastfoot
P\,G126.60-24.47 & 17.1239 &  38.2704 & 6.47 &$247(+69,-57)$&$633(+139,-116)$&$1238(+364,-266)$&$412(+353,-197)$&  1.54 &  1.96 &  3.01 & \\
P\,G293.78-69.78 & 17.4580 & -47.0360 & 5.97 &$324(+68,-57)$&$728(+142,-116)$&$1215(+302,-236)$&$220(+278,-109)$&  2.57 &  1.67 &  5.53 & L,S,C \\
P\,G132.16-47.01 & 19.3765 &  15.3889 & 5.10 &$293(+121,-80)$&$875(+322,-196)$&$1892(+890,-498)$&$480(+613,-237)$&  1.35 &  2.16 &  3.94 & \\
P\,G138.41-61.22 & 20.2445 &   0.7600 & 5.19 &$132(+55,-40)$&$371(+116,-83)$&$735(+267,-175)$&$93(+225,-46)$&  3.40 &  1.98 &  7.90 & \\
PE\,G137.77-41.86 &24.5465 &  19.6534 & 5.25 &$107(+59,-34)$&$296(+156,-76)$&$591(+400,-163)$&$129(+360,-64)$&  1.67 &  2.00 &  4.59 & \\
PE\,G138.84-31.60 &28.3941 &  29.3364 & 5.04 &$101(+50,-32)$&$315(+109,-76)$&$649(+273,-170)$&$44(+173,-22)$&  2.48 &  2.06 & 14.70 & \\
PE\,G299.42-35.45 &31.7309 & -81.1507 & 7.41 &$220(+50,-39)$&$487(+124,-90)$&$756(+264,-177)$&$48(+138,-24)$&  1.29 &  1.55 & 15.68 & \\
PE\,G145.18-33.98 &33.1898 &  25.3459 & 5.09 &$237(+102,-66)$&$793(+256,-181)$&$1822(+739,-467)$&$327(+573,-162)$&  1.62 &  2.30 &  5.58 & \\
\hline
\end{longtable}
\end{center}

\newpage
\begin{center}
\tabcolsep=0.11cm
\begin{longtable}{lcccccccccc}
\caption{MTXF photometry of candidate strongly lensed galaxies selected at 353\,GHz. Units and notes as in Table~\ref{tab:MTXFcandidates545GHz}. }
\label{tab:MTXFcandidates353GHz}\\
\hline
  \multicolumn{1}{c}{\textit{Planck} name} &
  \multicolumn{1}{c}{RA} &
  \multicolumn{1}{c}{DEC} &
  \multicolumn{1}{c}{$S_{217}$}&
  \multicolumn{1}{c}{$S_{353}$}&
  \multicolumn{1}{c}{$S_{545}$}&
  \multicolumn{1}{c}{$S_{857}$}&
  \multicolumn{1}{c}{SN$_{353}$} &
  \multicolumn{1}{c}{$S_{857}/S_{545}$}&
  \multicolumn{1}{c}{$S_{545}/S_{353}$}&
  \multicolumn{1}{c}{Note} \\
\hline
\endfirsthead

\multicolumn{11}{c}%
{{\bfseries \tablename\ \thetable{} -- continued from previous page}} \\
\hline
  \multicolumn{1}{c}{\textit{Planck} name} &
  \multicolumn{1}{c}{RA} &
  \multicolumn{1}{c}{DEC} &
  \multicolumn{1}{c}{$S_{217}$}&
  \multicolumn{1}{c}{$S_{353}$}&
  \multicolumn{1}{c}{$S_{545}$}&
  \multicolumn{1}{c}{$S_{857}$}&
  \multicolumn{1}{c}{SN$_{353}$} &
  \multicolumn{1}{m{1.2cm}}{$S_{857}/S_{545}$}&
  \multicolumn{1}{m{1.2cm}}{$S_{545}/S_{353}$}&
  \multicolumn{1}{c}{Note} \\
\hline
\endhead

\hline \multicolumn{11}{r}{{Continued on next page}} \\ \hline
\endfoot

\hline \hline
\endlastfoot
P\,G345.90-77.07 &   1.9719 & -36.1645 & $ 50\pm 30$ & $ 202\pm  58$ & $ 427\pm 110$ & $ 549\pm 195$ & 3.5 &  1.29 &  2.12 & H \\
P\,G314.57-62.41 &   3.7358 & -53.8762 & $ 75\pm 31$ & $ 287\pm  65$ & $ 422\pm  89$ & $ 577\pm 147$ & 4.4 &  1.37 &  1.47 & \\
P\,G123.48-41.23 &  13.3077 &  21.6338 & $ 66\pm 33$ & $ 374\pm  89$ & $ 432\pm 114$ & $ 942\pm 263$ & 4.2 &  2.18 &  1.16 & \\
P\,G127.01-82.91 &  13.3958 & -20.0593 & $ 66\pm 28$ & $ 255\pm  68$ & $ 290\pm 115$ & $ 653\pm 277$ & 3.7 &  2.25 &  1.14 & \\
P\,G297.94-67.76 &  15.7465 & -49.2554 & $ 54\pm 28$ & $ 334\pm  69$ & $ 294\pm  89$ & $ 271\pm 156$ & 4.9 &  0.92 &  0.88 & C \\
P\,G215.84-85.99 &  17.3622 & -27.2603 & $ 30\pm 32$ & $ 255\pm  70$ & $ 322\pm 103$ & $ 425\pm 223$ & 3.6 &  1.32 &  1.26 & \\
P\,G159.72-82.94 &  17.3930 & -21.4059 & $ 46\pm 32$ & $ 228\pm  72$ & $ 306\pm 111$ & $ 626\pm 246$ & 3.2 &  2.04 &  1.34 & \\
P\,G266.70-81.83 &  18.6412 & -33.5940 & $ 59\pm 31$ & $ 243\pm  73$ & $ 380\pm 113$ & $ 772\pm 195$ & 3.3 &  2.03 &  1.56 & \\
\hline
\end{longtable}
\end{center}
\end{document}